\documentclass[10pt,a4paper]{article} 

\usepackage[T1]{fontenc}
\usepackage{amsmath,mathtools,amssymb}
\usepackage{graphicx,relsize}
\usepackage{times}
\usepackage{multirow,lineno}
\usepackage{setspace,xcolor}
\usepackage{titlesec,etoolbox}
\usepackage{siunitx,booktabs,tabularx,ragged2e,threeparttable}
\usepackage[dvips]{xy}
\usepackage{microtype} 
\usepackage{setspace} 

\usepackage{times}
\usepackage[numbers,sort&compress]{natbib}  
\usepackage[colorlinks=true,
linkcolor=blue,
citecolor=blue,
urlcolor=magenta]{hyperref}

\usepackage[left=2.cm,right=2.cm,top=2.cm,bottom=2.cm]{geometry}

\selectfont

\titleformat{\section}
{\fontsize{10}{10}\selectfont\bfseries\raggedright} 
{\thesection.} 
{0.3em} 
{}
\titlespacing*{\section}{0pt}{10pt}{6pt}

\titleformat{\subsection}
{\fontsize{10}{10}\selectfont\bfseries\raggedright} 
{\thesubsection.}
{0.3em}
{}
\titlespacing*{\subsection}{0pt}{10pt}{6pt}

\titleformat{\subsubsection}
{\fontsize{10}{10}\selectfont\bfseries\itshape\raggedright} 
{\thesubsubsection.}
{0.3em}
{}
\titlespacing*{\subsubsection}{0pt}{10pt}{3pt}
\title{Self-organized biodiversity and species abundance distribution patterns in ecosystems with higher-order interactions}
\author{
    Ju Kang\textsuperscript{1}, 
	Yiyuan Niu\textsuperscript{2}, 
	Yuanzhi Li\textsuperscript{1},
    Chengjin Chu\textsuperscript{1,*}
}
\date{}
\newcommand{\affil}[2]{%
	\textsuperscript{#1}#2%
}

\makeatletter
\renewcommand{\maketitle}{
	\begin{center}
		{\LARGE\bfseries \@title\par}
		\vspace{1em}  
		{\normalsize \@author\par}
		\vspace{-0.3em} 
		\setlength{\parskip}{1pt}
		\@date
	\end{center}
}
\makeatother
\begin{document}
	
	\maketitle
	
	\begin{center}
		\affil{1}{School of Ecology, Sun Yat-sen University, Shenzhen 518107, China}\\
		\affil{2}{School of Physics, Sun Yat-sen University, Guangzhou 510275, China}\\
		\affil{*}{Corresponding author: \href{mailto:chuchjin@mail.sysu.edu.cn}{chuchjin@mail.sysu.edu.cn}}
	\end{center}	
	\begin{abstract}
		Explaining the emergence of self-organized biodiversity and species abundance distribution patterns remians a fundamental challenge in ecology. 
		While classical frameworks, such as neutral theory and models based on pairwise species interactions, 
		have provided valuable insights, they often neglect higher-order interactions (HOIs), whose role in stabilizing ecological communities is increasingly recognized. 
		Here, we extend the Generalized Lotka-Volterra framework to incorporate HOIs and demonstrate that 
		these interactions can enhance ecosystem stability and prevent collapse.
		Our model exhibits a diverse range of emergent dynamics, including self-sustained oscillations, 
		quasi-periodic (torus) trajectories, and intermittent chaos. 
		Remarkably, it also reproduces empirical species abundance distributions observed across diverse natural communities. 
		These results underscore the critical role of HOIs in structuring biodiversity and offer a broadly applicable theoretical 
		framework for capturing complexity in ecological systems.
		\vspace{0.3cm}		
	
	\end{abstract}	
	\section{Introduction}
	A central challenge in ecology is explaining the vast biodiversity observed across the planet, 
	ranging from the macro to microbial scale~\cite{PennisiE2005,DanielR2005,HoornC2010,deVargasC2015,DonghaoWu2025}. 
	Natural ecosystems are astonishingly rich in species: tropical forests support the coexistence of thousands of plant and 
	vertebrate species, a single gram of soil may harbor between 2,000 and 18,000 microbial taxa~\cite{DanielR2005,HoornC2010}, 
	and the photic zone of the global ocean contains an estimated 150,000 eukaryotic plankton species~\cite{deVargasC2015}. 
	The persistence of such extraordinary biodiversity has remained a fundamental enigma in ecology~\cite{PennisiE2005}.
	This challenge arises from the inherent nonlinearity and complexity of ecological systems. 
	Species interact not only with the environment but also with each other in intricate ways, resulting in feedback loops 
	that may either promote or hinder coexistence. 
	Traditionally, ecological models have addressed these dynamics by assuming that species interact exclusively in pairwise fashion, 
	and that the behavior of an entire community can be derived from the sum of all pairwise interactions~\cite{LotkaAJ1910,VolterraV1928,JonathanMLevine2017}. 
	However, this simplifying assumption often fails to capture the emergent complexity and coexistence patterns 
	observed in real ecosystems~\cite{NathanJB2015,JJelleLever2020}.
	
	In particular, when the interaction between two species is modulated by the presence of a third, 
	pairwise frameworks become insufficient. 
	These more complex interactions, termed higher-order interactions (HOIs), involve three or more species and 
	have long been acknowledged in ecology~\cite{PeterAbrams1980,MarkJPomerantz1981,PeterAbrams1983,AdlerFR1994,WadeBWorthen1991,IanBillick1994,EyalBairey2016,FedericoBattiston2020,JacopoGrilli2017,TheoGibbs2022,PragyaSingh2021,MargaretMMayfield2017,TheoLGibbs2024,EricDKelsic2015,JohnVandermeer2024,MiltonBarbosa2023,HarryMickalide2019,YuanzhiLi2021,YinglinLi2020}. 
	Yet, their role in supporting self-organized coexistence and shaping species abundance distributions remains theoretically underexplored.
	Since the seminal work of Billick and Case~\cite{IanBillick1994}, who formally defined the concept of HOIs, 
	a growing number of theoretical and empirical studies have sought to uncover their ecological implications. 
	These include models of competition~\cite{JacopoGrilli2017,TheoGibbs2022,TheoLGibbs2024}, 
	hypernetwork representations~\cite{FedericoBattiston2020}, random community frameworks~\cite{PragyaSingh2021}, 
	oscillator-based models~\cite{JohnVandermeer2024}, and generalized statistical approaches such as generalised linear model and collective competition theories~\cite{AdlerFR1994,YinglinLi2020}. 
	Complementary experimental work has identified HOIs in microbial consortia~\cite{EyalBairey2016,EricDKelsic2015,HarryMickalide2019}, 
	plant and animal communities~\cite{WadeBWorthen1991,MiltonBarbosa2023,YuanzhiLi2021}, and network-level analyses of ecosystem 
	interactions~\cite{MargaretMMayfield2017,TheoLGibbs2024}. 
	These studies collectively suggest that HOIs can play a key role in promoting community stability and biodiversity. 
	However, the precise mechanisms by which HOIs enable self-organized biodiversity in multi-species ecosystems remain poorly 
	understood~\cite{PennisiE2005}.
	
	In this study, we extend the classical Generalized Lotka-Volterra (GLV) framework to incorporate HOIs
	and demonstrate that HOIs can support the stable coexistence of large and diverse populations through self-organization. 
	Depending on system parameters, the model exhibits a rich spectrum of dynamical behaviors, including steady-state equilibria, 
	periodic, multi-periodic, and quasi-periodic oscillations, and chaotic dynamics. 
	Crucially, we show that this framework not only prevents ecosystem collapse but also quantitatively reproduces empirical 
	rank-abundance distributions observed across a wide range of real-world ecological communities,
	including wild bee communities from meadows and deserts in North America~\cite{MichaelRoswell2021,HubbellSP2001}, 
	insect populations from U.S. Long-Term Ecological Research sites~\cite{MichaelSCrossley2020}, 
	bird communities from global datasets~\cite{HubbellSP2001,Bird6,Bird7,Bird8,Bird9,Bird11,Bird15,Bird16,Bird17,Bird18,Bird23,Bird30,Bird34,Bird36,Bird37,Bird38,Bird51}, 
	bat populations from the tropical forests of Mexico and Trinidad~\cite{EstradaA2001,ClarkeFM2005}, 
	planktonic species from the Norwegian Sea and Antarctic regions~\cite{FuhrmanJA2008}, 
	and controlled bacterial consortia from laboratory experiments~\cite{Hu2022}.
	\section{Results}
	\subsection{Theoretical framework}
\begin{figure}[h]
		\centering
		\includegraphics[width=17cm]{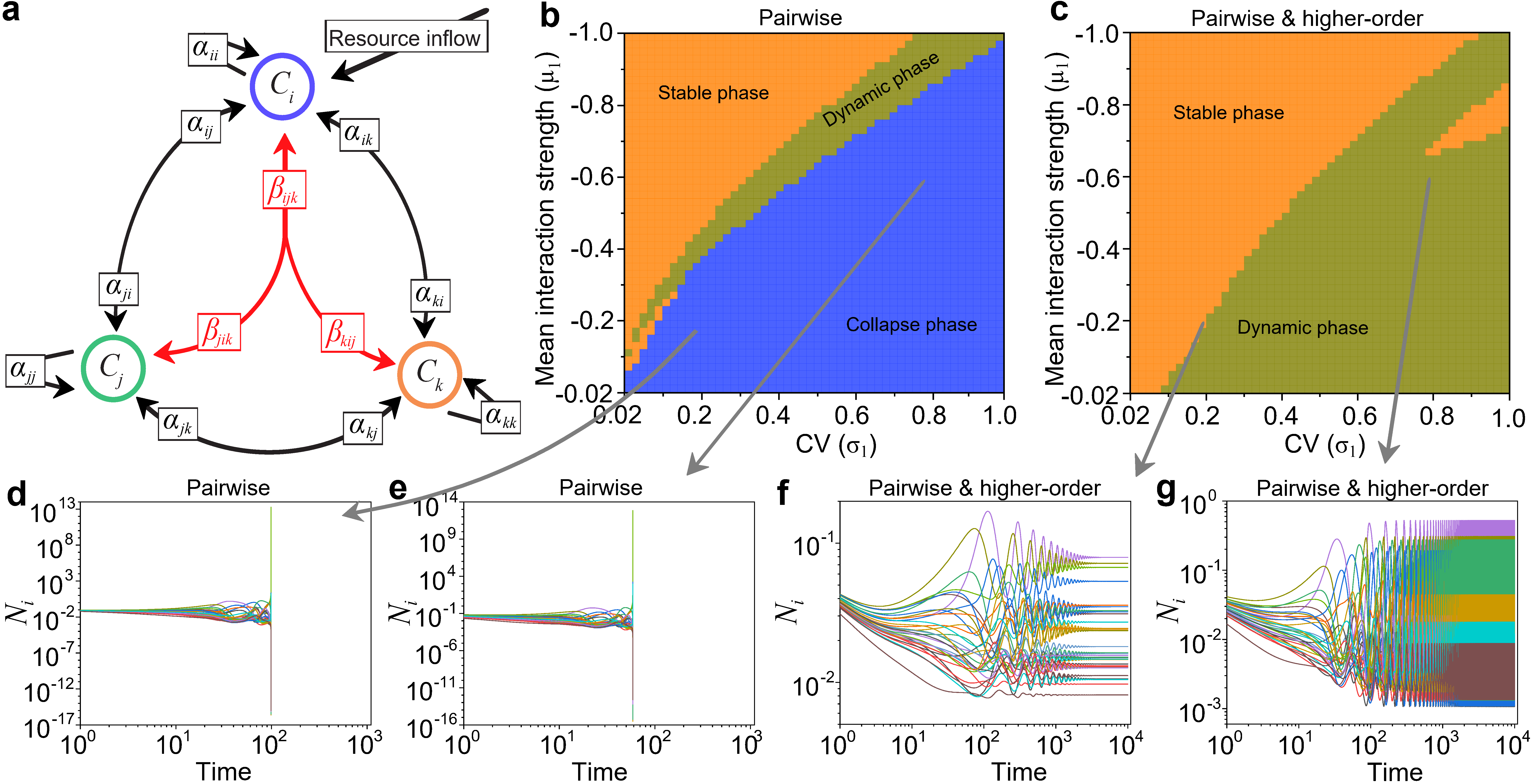}
		\caption{\label{pairwise higher-order} 
		Higher-order interactions prevent ecosystems collapse.
		(a) Schematic representation of the generalized model incorporating both pairwise and higher-order interactions among $S$ consumer species. 
		(b) Species collapse occurs when dynamics are governed solely by pairwise interactions.
		(c) The introduction of higher-order interactions stabilizes the system, enabling persistent coexistence through self-organized dynamics, even under the same initial conditions as in (b).
		(d-g) Representative time series of species abundances from simulations with $S = 32$ species, illustrating the emergence of diverse dynamical regimes.
		For full simulation details, see SM Sec. \ref{Simulation details}.	}
	\end{figure}

	We consider an ecological community of $S$ species, where population dynamics are governed by both pairwise interactions and HOIs 
	(see Fig.~\ref{pairwise higher-order}a). 
	To facilitate the emerengence of HOIs, we introduce a small dispersal rate $d_{i}$ for each species~\cite{Hu2022}. 
	Under these conditions, the population dynamics can be described by the GLV model as follows:
	\begin{equation}
	\frac{d{N}_i}{dt} = N_i \left( r_i + \sum_{j=1}^S \alpha_{ij}N_j + \sum_{j,k=1}^S \beta_{ijk}N_jN_k \right) + d_i.
		\label{GLV}
	\end{equation}
	Let ${N}_{i}~(i,j,k=1,\dots,S)$ denote the abundances of species $i$. 
	The pairwise interaction structure is captured by the matrix 
	$\textbf{A}=(\alpha_{ij})\in\mathbb{R}^{S\times S}$, in which $\alpha_{ij}$ quantifies the average effect of species $j $
	on the per capita growth rate of species $i$. 
	Higher-order interactions are encoded by the third-order tensor 
	$\textbf{B}=(\beta_{ijk})\in\mathbb{R}^{S\times S^{2}}$, where $\beta_{ijk}$ characterizes the influence of 
	the joint presence of species $j$ and $k$ on species $i$.
	Together, matrices $\textbf{A}$ and $\textbf{B}$ define the expected ecological interaction structure of the community, 
	averaged over environmental fluctuations and spatial heterogeneity.
	Negative values of $\alpha_{ij}<0,~\beta_{ijk}<0$ indicate competitive interactions, 
	whereas positive values of $\alpha_{ij}>0,~\beta_{ijk}>0$ indicate cooperative interactions. 
	This formulation allows for communities governed primarily by competition  ($\alpha_{ij}<0,~\beta_{ijk}<0$), 
	cooperation  ($\alpha_{ij}>0,~\beta_{ijk}>0$), or any mixture of the two.
	The effective growth rate $r_{i}$ represents the net per capita growth of species $i$, encompassing intrinsic biological traits, 
	abiotic environmental influences, (e.g., temperature, nutrients), and the impact of unmodeled biotic interactions~\cite{Deng2022, Arnoldi2022,Deng2024}.
	In our simulations, the entries of pairwise interaction matrix $\textbf{A}$ and the higher-order interaction tensor $\textbf{B}$ 
	are drawn independently from a Gaussian distributions: $\textbf{A}=\mathcal{N}(\mu_{1}, \sigma_{1})$ 
	and $\textbf{B}=\mathcal{N}(\mu_{2}, \sigma_{2})$), where $\mu$ and $\sigma$ represent the mean and standard deviation of
	the interaction coefficients.
	\subsection{Stability analysis}
	To assess the stability of the coexistent equilibrium $N_i^*$~($i=1,\cdots,S$) 
	(The existence of $N_i^*$ can be seen in Supplementary Material (SM) Sec.~\ref{coexistence equilibrium} 
	for details), we analyse the Jacobian matrix derived from both pairwise and HOIs
	(see Methods~\ref{Methods Local} and SM Sec.~\ref{Jacobian} for details). 
	The Jacobian encodes the local sensitivities of each species' abundance to small perturbations 
	in the abundances of all species, including self-effects.
	Local stability is determined by evaluating the Jacobian at the coexistent equilibrium 
	and examining its eigenvalues: the equilibrium is locally stable if all eigenvalues have negative 
	real parts. To assess global stability, we construct a relative entropy function that serves as 
	a Lyapunov function for the system~(\ref{GLV}) 
	(see Methods~\ref{Methods Global} and SM Sec.~\ref{global stability} for details).
	
	This numerical example demonstrates the stability properties of the coexistent equilibrium $N_i^*$ 
	in system~(\ref{GLV}). For computational tractability, we consider a five-species community (i.e., $S=5$). 
	As illustrated in Fig.~\ref{stability}, the mean strength of HOIs 
	plays a critical role in shaping the system's dynamical behavior. 
	When the mean HOI strength is set to $\mu_{2} = -0.25$, system~(\ref{GLV}) converges to a stable equilibrium (see Fig.~\ref{stability}a-b). 
	The corresponding equilibrium is $N^* = (0.0539,0.1193,0.0211,0.2110,0.4942)$, 
	with the Jacobian eigenvalues given by $\lambda_{1,2}=-0.0177 \pm 0.0175i$, $\lambda_{3,4}=-0.0632 \pm 0.0348i$, and $\lambda_{5}=-0.5434$. 
	Since all eigenvalues have negative real parts, the equilibrium is locally asymptotically stable (see Fig.~\ref{stability}c). 
	Furthermore, a small perturbation around  $N^*$	yields a Lyapunov derivative of $\frac{dV}{dt} = -7.59\times 10^{-5}<0$,
	confirming the global stability of the equilibrium. 
	In contrast, when the mean HOI strength is decresed to $\mu_{2} = -0.15$, the system exhibits a stable limit cycle (see Fig.~\ref{stability}d-e). 
	The new equilibrium is $N^* = (0.0248,0.0341,0.0175,0.3677,0.6885)$, with Jacobian eigenvalues: $\lambda_{1,2}=-0.0615 \pm 0.0030i$, 
	 $\lambda_{3}=-0.6737$, $\lambda_{4}=-0.1969$, $\lambda_{5}=0.0096$. 
	The presence of a positive eigenvalue (highlighted as a red dot in Fig.~\ref{stability}f) indicates that the equilibrium is unstable. 
	However, the Lyapunov derivative remains negative, $\frac{dV}{dt} = -7.50\times 10^{-5}<0$, demonstrating convergence to a globally stable limit cycle.
	
	\subsection{HOIs prevent ecosystems collapse}
	To investigate mechanisms that prevent ecosystem collapse caused by pairwise interactions, 
	we consider a generic community of $S$ species subject to both pairwise and HOIs.
	In the absence of HOIs, system~(\ref{GLV}) collapses across a broad region of parameter space (see Fig.~\ref{pairwise higher-order}b),
	with long-term coexistence failing to emerge (see Fig.~\ref{pairwise higher-order}d-e). 
	These results confirm that pairwise interactions alone are insufficient to sustain species-rich communities under realistic conditions.
	By introducing HOIs into the system, we observe a striking shift in dynamics: the community exhibits stable 
	or oscillatory coexistence across the same parameter regimes (see Fig.~\ref{pairwise higher-order}c).
	This shift arises from the modulatory effect of HOIs, which reshape the effective interaction landscape. 
	Specifically, HOIs can stabilize communities or induce persistent population cycles in scenarios where 
	pairwise dynamics would otherwise lead to collapse (compare Fig.~\ref{pairwise higher-order}f-g with Fig.~\ref{pairwise higher-order}d-e).
	
	To further investigate the role of HOIs in promoting species coexistence, 
	we compute the fraction of coexisting species in each pixel across 60 independent simulations with randomized interaction matrices 
	(see Fig.~\ref{randfig}a-b). 
	In both scenarios-pairwise interactions alone and pairwise interactions with HOIs-stochasticity substantially affects coexistence 
	outcomes, as random seeds alter the structure of the pairwise matrix $\textbf{A}$ and the HOI tensor $\textbf{B}$.
	Despite this stochastic variation, a large collapse region persists when only pairwise interactions are present (see Fig.~\ref{randfig}a-b). 
	In contrast, the inclusion of HOIs consistently facilitates higher levels of coexistence across diverse conditions.
	Moreover, we observe a negative relationship between the coefficient of variation $\sigma_{1}$ 
	of the pairwise interaction matrix and the mean high-order interaction strength required to support coexistence (see Fig.~\ref{randfig}c-d). 
	This implies that even relatively weak HOIs can rescue systems from collapse when pairwise interaction variability is high,
	 enabling self-organization and robust species coexistence.
	\begin{figure}[h]
		\centering
		\includegraphics[width=15cm]{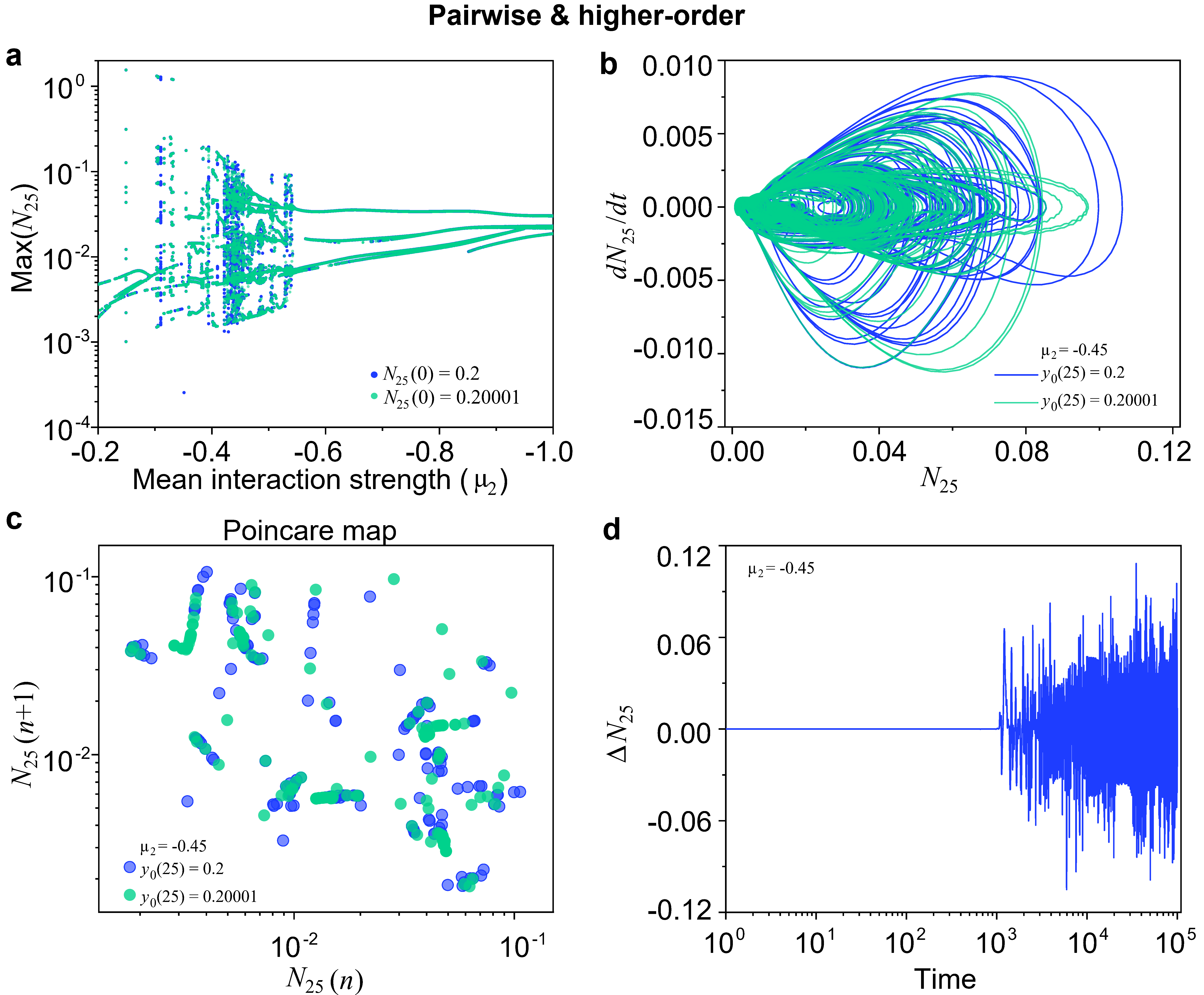}
		\caption{\label{Bifurcation-mu2} 
		Emergence of chaotic dynamics induced by pairwise and higher-order interactions.
		 (a-c) The blue and green dots represent the simulation results for initial conditions of $N_{25}(0)=0.2$ and $N_{25}(0)=0.20001$, 
		 respectively, with all other initial values set to $N_{i}(0)=0.2$ ($i=1,\cdots,32 $). 
		 (a) Bifurcation diagram of the system~(\ref{GLV})  as a function of the mean higher-order interaction strength $\mu_2$. 
		 (b-c) Representative chaotic dynamics at $\mu_2=-0.45$, consistent with the bifurcation structure in panel (a),
		 shown via a two-dimensional phase space projection and the corresponding Poincaré map.
		 (d) Sensitivity analysis of system~(\ref{GLV}) under the same parameters as in panels (b-c). 
		 The divergence $\Delta N_{25}$ quantifies the sensitivity to initial conditions, indicative of chaos.
		 All simulations were performed with $S = 32$ species. See SM Sec. \ref{Simulation details} for simulation details.}	
	\end{figure}
	\subsection{Emergence of chaotic dynamics from pairwise interactions and HOIs}

	In the 1970s, May~\cite{MayRM1974,MayRM1976} demonstrated that simple population models can exhibit complex chaotic dynamics. 
	Since then, a wide range of theoretical studies have shown that chaos can emerge from various ecological mechanisms, 
	including resource competition~\cite{JefHuisman1999,JefHuisman2006},  predator-prey interactions~\cite{MichaelEGilpin1976,JohnVandermeer1993}, 
	and trophic cascades in food chains~\cite{AlanHastings1991,EgbertHvanNes2004}. 
	Empirical evidence has also confirmed the presence of chaos in ecological systems. 
	Long-term observations of plankton communities have documented chaotic dynamics in nature~\cite{Benincà2008}, 
	while short-term laboratory experiments and artificial ecosystems have exhibited similar behaviors~\cite{RFCostantino1997,LutzBecks2005}.
	Here, we show that HOIs among multiple species can also give rise to chaos 
	(see Figs.~\ref{Bifurcation-mu2}, \ref{SF-Bifurcation-mu2}, \ref{SF-Bifurcation-mu2-phase}, \ref{bif_Poincare}, 
	\ref{bifurcation_SM00}, \ref{bif_Poincare_SM}). 
	This finding provides a theoretical foundation for understanding chaotic dynamics observed in real-world ecosystems, 
	especially those shaped by complex multi-species interactions.

	Figs.~\ref{Bifurcation-mu2}a,~\ref{SF-Bifurcation-mu2},~\ref{bif_Poincare}a,~\ref{bifurcation_SM00}, and~\ref{bif_Poincare_SM}a 
	reveal a cascade of non-periodic fluctuations in species abundances, 
	indicating a dynamical transition from non-chaotic to chaotic states as the parameters $\mu_{2}$ or $r_{i}$ vary. 
	Notably, system~(\ref{GLV}) exhibits period-3 orbits (see Fig.~\ref{bif_Poincare}a), which, 
	according to the theorem of Li and Yorke~\cite{TienYienLi1975}, imply the existence of chaos.
	For example, when $\mu_{2} = -0.45$ or $r = 0.8$, the system exhibits sustained chaotic dynamics, 
	as demonstrated in Figs.~\ref{Bifurcation-mu2}b,~\ref{SF-Bifurcation-mu2-phase},~\ref{bif_Poincare}b, and~\ref{bif_Poincare_SM}b.
 
	To further validate the presence of chaos, we conducted a sensitivity analysis by initiating the system (\ref{GLV}) with 
	two nearly identical initial conditions. 
	Specifically, simulations were run with $N_{25}(0) = 0.2$ (blue) and $N_{25}(0) = 0.20001$ (green), 
	while all other species were initialized at $0.2$. Despite a minute difference of only $10^{-5}$, 
	the resulting trajectories diverge markedly over time.
	This divergence is clearly illustrated across bifurcation diagrams, phase portraits, Poincaré maps, and time series comparisons 
	(see Figs.~\ref{Bifurcation-mu2}a-d and \ref{bif_Poincare}a-d). 
	Such pronounced sensitivity to initial conditions is a defining characteristic of chaotic dynamics in system~(\ref{GLV}).
	
	We further examine a generic scenario in which multiple consumer species compete for resources exclusively through HOIs, 
	with no pairwise interactions present ($\alpha_{ij} = 0$). 
	In this HOI-only system, population dynamics are governed entirely by Eq.~(\ref{GLV}).
	Remarkably, even in the absence of pairwise terms, the community can self-organize into a diverse and stable structure, 
	supporting the coexistence of multiple species in three distinct dynamical regimes: 
	steady-state equilibrium, periodic oscillations, and quasi-periodic oscillations (see Fig.~\ref{self-organized dynamics}). 
	As shown in Fig.~\ref{self-organized dynamics}c, e-f, the system converges to a stable invariant torus, 
	a hallmark of quasi-periodic dynamics driven solely by HOIs.
	Interestingly, transitions between these regimes are modulated by the mean strength of HOIs. 
	As the average interaction strength increases, the system shifts from oscillatory behavior toward stable coexistence 
	(see Fig.~\ref{self-organized dynamics}).
	
	\begin{figure}[ht!]
	\centering
	\includegraphics[width=15cm]{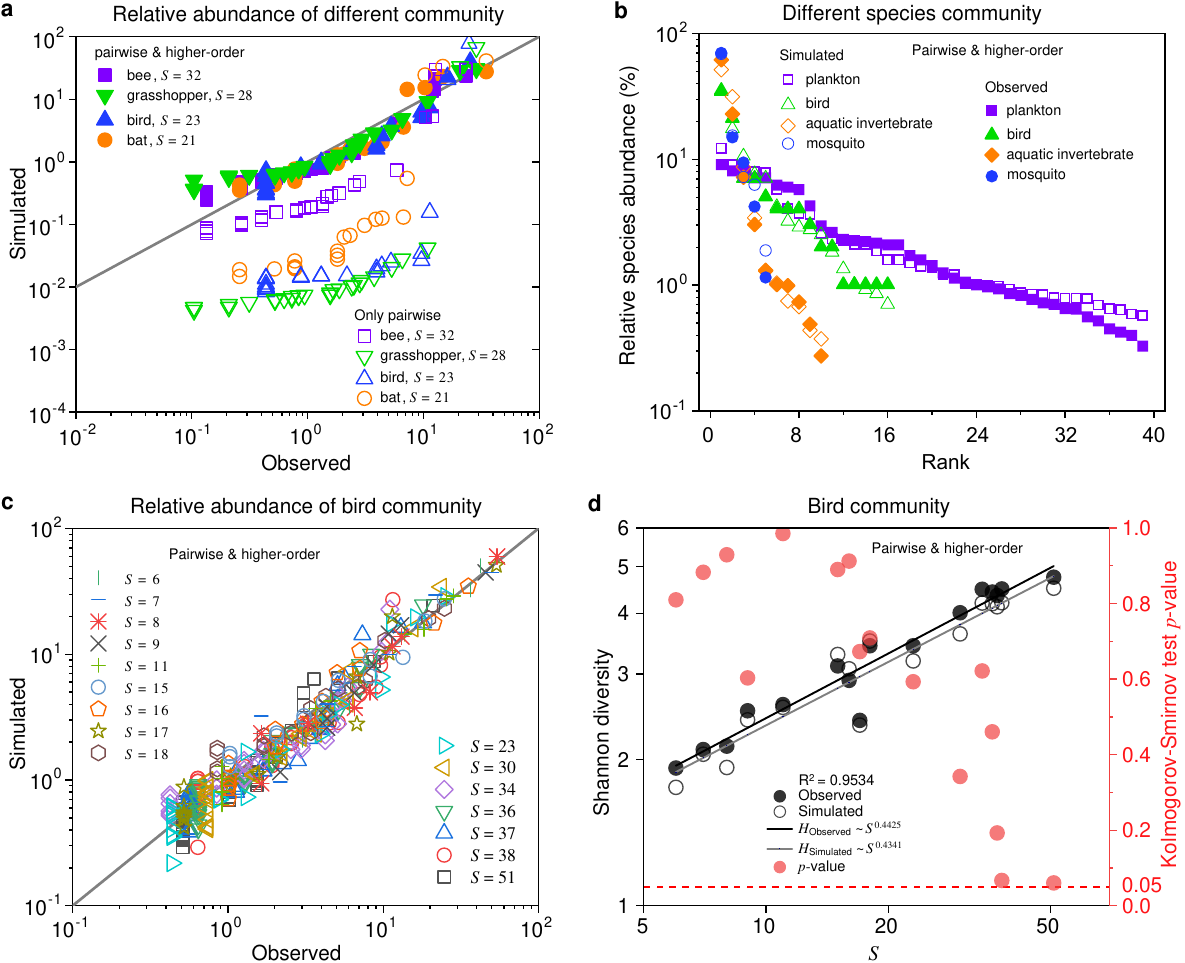}
	\caption{\label{FigRank} 
	Higher-order interactions shape species distribution patterns across ecological communities.
		(a) Comparison of species abundance distributions across communities. Hollow markers represent simulations with 
		pairwise interactions only; solid markers incorporate both pairwise and HOIs.
		Observed data are from published empirical studies~\cite{MichaelRoswell2021,MichaelSCrossley2020,Bird23,EstradaA2001}.
		(b) HOIs reproduce the characteristic S-shaped rank-abundance curves observed in empirical ecosystems~\cite{HubbellSP2001,FuhrmanJA2008,MichaelSCrossley2020}.
		Solid markers denote empirical data; hollow markers show corresponding simulation outcomes.
		(c) Direct visual comparison of species abundance distributions in a bird community. Empirical observations are sourced from 
		Cornell Lab: Birds of the World~\cite{Bird6,Bird7,Bird8,Bird9,Bird11,Bird15,Bird16,Bird17,Bird18,Bird23,Bird30,Bird34,Bird36,Bird37,Bird38,Bird51}.
		(d) Comparison of Shannon diversity indices and Kolmogorov-Smirnov (K-S) test \emph{p}-values quantifies the similarity 
		between observed and simulated distributions in (c). 
		At the 0.05 significance threshold, none of the \emph{p}-values indicate statistically significant differences.
		(a-d) All simulations were evaluated at time $t = 1.0 \times 10^{5}$. See SM Sec.\ref{Simulation details} for full details.}	
	\end{figure}	
	\subsection{Species abundance distribution patterns across diverse ecological communities}
	The quantitative explanation of biodiversity is often captured through species abundance distributions. 
	Several theoretical frameworks, such as neutral theory~\cite{HubbellSP2001}, metabolic trade-offs~\cite{AnnaPosfai2017}, 
	and intraspecific interference among predators~\cite{JuKangeLife,JuKangCSF,JuKangQB}, 
	have successfully reproduced the characteristic rank-abundance curves observed in ecological communities. 
	However, these models typically neglect HOIs, thereby limiting their ability to capture the full complexity of species coexistence.
	Interestingly, despite differences in community composition and ecosystem type, empirical species abundance curves 
	often exhibit strikingly similar shapes across taxa~\cite{HubbellSP2001,AnnaPosfai2017,JuKangeLife,JuKangCSF,JuKangQB}. 
	This raises a fundamental question: To what extent do HOIs shape these ubiquitous distribution patterns?
	To address this, we compiled empirical data from a wide range of ecosystems-including bird, bat, bee, insect, bacterial, 
	and plankton communities~\cite{Hu2022,MichaelRoswell2021,HubbellSP2001,FuhrmanJA2008,MichaelSCrossley2020,EstradaA2001,ClarkeFM2005,Bird6,Bird7,Bird8,Bird9,Bird11,Bird15,Bird16,Bird17,Bird18,Bird23,Bird30,Bird34,Bird36,Bird37,Bird38,Bird51}. 
	We then simulated a well-mixed ecological community governed by the GLV model that explicitly incorporates HOIs (see Eq.~(\ref{GLV})). 
	In these simulations, species interaction coefficients were drawn from a Gaussian distribution with a coefficient of variation (CV) 
	of approximately 0.3.
	
	Figs.~\ref{FigRank},~\ref{elife-2020bee-2001batdata},~\ref{Hudata},~\ref{NEE_mosquito-acquatic}, 
	and~\ref{NEE-Stream insects-Sevilleta} compare simulated species distribution patterns with empirical data from a wide range of 
	ecological communities. 
	The simulations, based on ordinary differential equation (ODE) models incorporating HOIs, 
	closely reproduce the rank-abundance curves observed in nature (see Fig.~\ref{FigRank}b).
	While minor discrepancies appear for species with low relative abundance in certain ecosystems 
	(see Fig.~\ref{FigRank}a, d,~\ref{elife-2020bee-2001batdata}a-c,~\ref{Hudata}a,~\ref{NEE_mosquito-acquatic}a, c,~
	\ref{NEE-Stream insects-Sevilleta}a, c), 
	these deviations likely reflect stochastic drift and sampling variability inherent in empirical surveys. 
	Importantly, the overall diversity patterns remain consistent: both simulated and observed communities yield 
	comparable Shannon entropy values (see Figs.~\ref{FigRank}d,~\ref{elife-2020bee-2001batdata}d-f,~\ref{Hudata}b,~\ref{NEE_mosquito-acquatic}b, d,~
	\ref{NEE-Stream insects-Sevilleta}b, d).
	To quantitatively assess the agreement, we applied the Kolmogorov-Smirnov (K-S) test at a 0.05 significance level.
	Across all cases, the test failed to reject the null hypothesis, indicating no statistically significant difference between 
	the simulated and observed species distributions
	(see Figs.~\ref{FigRank}d,~\ref{elife-2020bee-2001batdata}d-f ,~\ref{Hudata}b,~\ref{NEE_mosquito-acquatic}b, d,
	~\ref{NEE-Stream insects-Sevilleta}b, d).
	
	Interestingly, when only pairwise interactions were considered, a clear mismatch emerged between simulation results and 
	empirical species abundance distributions (see Fig.~\ref{FigRank}a). 
	By contrast, incorporating HOIs markedly improved the fit: all Kolmogorov-Smirnov (K-S) test \emph{p}-values exceeded the 0.05 threshold, indicating statistical consistency with observed patterns. These results highlight the essential role of HOIs in shaping community structure and suggest that models incorporating such interactions offer a powerful and generalizable framework for understanding the organization and dynamics of real-world ecosystems.
	
	\section{Discussion}
	Although numerous studies have shown that HOIs can enhance ecosystem stability, promote biodiversity, 
	and enrich our understanding of ecological dynamics~\cite{PeterAbrams1980,MarkJPomerantz1981,PeterAbrams1983,AdlerFR1994,WadeBWorthen1991,IanBillick1994,EyalBairey2016,FedericoBattiston2020,JacopoGrilli2017,TheoGibbs2022,PragyaSingh2021,MargaretMMayfield2017,TheoLGibbs2024,EricDKelsic2015,JohnVandermeer2024,MiltonBarbosa2023,HarryMickalide2019,YuanzhiLi2021,YinglinLi2020}, 
	their precise role in stabilizing or destabilizing species coexistence remains poorly understood, both empirically and theoretically~\cite{PeterAbrams1983,JacopoGrilli2017,DavidSloanWilson1992,AndrewDLetten2019}. 
	In particular, it remains unclear how HOIs among multiple species give rise to long-term coexistence 
	and explain the rank-abundance distributions observed in diverse ecological communities.
	To address this gap, we analyze a GLV model that incorporates both pairwise and HOIs. 
	Through a combination of analytical techniques and large-scale simulations, we demonstrate that HOIs can enable self-organized 
	coexistence among numerous species. 
	Our results not only reveal a generic mechanism for stabilizing biodiversity but also quantitatively reproduce the universal 
	rank-abundance curves observed across a wide range of ecosystems-including bird, bat, bee, insect, bacterial, 
	and plankton communities~\cite{Hu2022,MichaelRoswell2021,HubbellSP2001,FuhrmanJA2008,MichaelSCrossley2020,EstradaA2001,ClarkeFM2005,Bird6,Bird7,Bird8,Bird9,Bird11,Bird15,Bird16,Bird17,Bird18,Bird23,Bird30,Bird34,Bird36,Bird37,Bird38,Bird51}.

	The central focus of our study is to understand how HOIs mitigate ecosystem collapse induced by 
	pairwise interactions. 
	We find that increasing the mean strength of HOIs enhances species coexistence and drives a transition in population dynamics:
	from chaotic to non-chaotic regimes. 
	Likewise, when only HOIs are present, increasing their strength leads to a shift from self-organized oscillatory coexistence 
	to stable steady-state configurations. 
	These results suggest that stronger HOIs play a stabilizing role in ecological communities~\cite{EyalBairey2016,JacopoGrilli2017,TheoGibbs2022,PragyaSingh2021,MargaretMMayfield2017,TheoLGibbs2024,EricDKelsic2015,JohnVandermeer2024,MiltonBarbosa2023,HarryMickalide2019}.
	In contrast, the intrinsic growth rate $r_i$ exerts a more nuanced influence on community dynamics. 
	As $r_i$ increases, the system undergoes a dynamical transition: first from stable coexistence to oscillatory behavior, and ultimately to chaos. 
	This highlights the delicate interplay between intrinsic species traits and interaction structure in shaping long-term ecological stability.
	
	The distribution of species diversity has long been interpreted through the lens of neutral theory, which posits that all species 
	have equivalent competitive abilities~\cite{HubbellSP2001}. 
	More recently, mechanistic models incorporating intraspecific predator interference have successfully reproduced 
	the rank-abundance curves observed across a range of ecological communities~\cite{JuKangeLife,JuKangCSF,JuKangQB}. 
	However, these frameworks largely overlook higher-order interactions (HOIs), 
	which are increasingly recognized as key drivers of ecological dynamics. 
	In particular, HOIs are pervasive in microbial ecosystems, where species engage in complex,
	nonlinear interactions~\cite{EricDKelsic2015,HarryMickalide2019}.
	To address this gap, our study demonstrates that HOIs alone can quantitatively replicate the universal species abundance patterns 
	observed in diverse ecosystems. 
	By capturing interaction structures beyond the pairwise level, our model offers a generalizable framework for understanding 
	biodiversity across ecological scales. 
	While these findings advance our theoretical understanding, the mechanisms underlying the origin and maintenance of biodiversity 
	remain among the most fundamental open questions in ecology.
	\section{Conclusion}	
	Explaining self-organized biodiversity and the distribution of species abundance remains a central challenge in ecology. 
	While established frameworks-such as neutral theory~\cite{HubbellSP2001} 
	and models based on pairwise interactions~\cite{JuKangeLife,JuKangCSF,JuKangQB} have yielded valuable insights, 
	they often overlook the role of HOIs among species. 
	Here, we introduce a GLV model that incorporates HOIs and show that such interactions can prevent ecosystem collapse 
	by promoting dynamic complexity. 
	Specifically, we demonstrate that HOIs can drive transitions from periodic oscillation to chaotic dynamics, underscoring their importance in 
	regulating ecological stability. 
	When calibrated against empirical data, our model quantitatively reproduces species abundance distributions 
	across a wide range of ecosystems. 
	More broadly, it offers a flexible and unifying framework for understanding biodiversity in complex ecological communities.
	\section{Methods}
	\subsection{Simulation Details}
	To numerically solve the population dynamics, we used the ODE45 solver for ordinary differential equations (ODEs) in MATLAB R2018b. 
	The interaction parameters, including pairwise interactions ($\alpha_{ij}$) and higher-order interactions ($\beta_{ijk}$, 
	with $i,j,k = 1, \dots, S$), were sampled from normal distributions with specified means and standard deviations. 
	All species were initialized with an abundance of 0.2. 
	Simulations were conducted for $10^5$ time units using the default time step of ODE45. 
	A species was considered extinct if its abundance dropped below $10^{-14}$. 
	Conversely, if any species exhibited unbounded growth (abundance explosion), the system was classified as divergent, 
	indicating ecosystem collapse. 
	Throughout the main text, we report numerical results corresponding to coexisting equilibria, 
	which are guaranteed to exist under the dynamics of Eq.~(\ref{GLV}).
	\subsection{Local asymptotic stability}\label{Methods Local}
	To assess whether a given set of interactions leads to a locally stable coexistence equilibrium $N_i^*$ 
	(whose existence is discussed in SM Sec.~\ref{coexistence equilibrium}), we construct the Jacobian matrix of the system. 
	The Jacobian matrix, comprising the partial derivatives of each species' growth rate with respect to all species' abundances,
	quantifies the system's response to small perturbations near equilibrium. 
	We evaluate the Jacobian at the coexistent equilibrium and compute its eigenvalues to determine local stability. 
	Specifically, the $(i, j)$-th entry of the Jacobian matrix for system~(\ref{GLV}) at $N_i^*$ is given by 
	(see SM Sec.~\ref{Jacobian} for details):
	\begin{equation}
		J_{lm} = \delta_{lm} \left( -\frac{d_l}{N_l^*} \right) + N_l^* \left( \alpha_{lm} + 2\sum_{k=1}^S \beta_{lmk}N_k^* \right).
		\label{eq:Jacobian_simplified1}
	\end{equation}
	The local stability of the coexistent equilibrium $N_i^*$ is determined by the eigenvalues of the Jacobian matrix $J$
	evaluated at that point. 
	If all eigenvalues of $J$ have negative real parts, the equilibrium is locally asymptotically stable under system~(\ref{GLV}). 
	Conversely, if at least one eigenvalue has a positive real part, the equilibrium $N_i^*$ is locally unstable.
	\subsection{Global asymptotic stability}\label{Methods Global}
	To assess global stability of the coexistent equilibrium, we construct an appropriate Lyapunov function, 
	as detailed in SM Sec.~\ref{global stability}. 
	\begin{equation}
		V(\mathbf{N}) = \sum_{i=1}^n \left( N_i - N_i^* - N_i^* \ln \frac{N_i}{N_i^*} \right).
		\label{Lyapunov_function}
	\end{equation}
	If the time derivative of $V$ is less than zero (i.e. $\frac{dV}{dt}<0$), then system~(\ref{GLV}) is globally stable at the 
	coexistent equilibrium $N_i^*$. Otherwise, it is unstable.	   
    
	\section*{Compliance with Ethics requirements}
    This article does not contain any studies with human or animal subjects.
    	
	\section*{Acknowledgments}
	This work was supported by the National Natural Science Foundation of China (Nos.32330064 and 32525006) and the China National Key Research Development Program (No.2022YFF0802300). 
	
	\section*{Conflict of interest}
	The authors declare that they have no known competing financial interests or personal relationships that could have appeared to influence the work reported in this paper.
	
	\section*{Author contributions}
	J.K. and C.C. conceived the project and planned the study. All authors supervised the study, developed the model, analyzed the data, carried out the theoretical analysis and numerical simulations, analyzed the simulation results, and wrote the paper. 
	
	\section*{Data and code availability}
	All data and code used in this paper are available upon request to the corresponding authors.
	
    	
	
		\newpage
		\textbf{\LARGE{Supplementary Materials}}
		
		\renewcommand \thesection {\Roman{section}}
		\renewcommand \thesubsection {\Alph{subsection}}
		\renewcommand \thesubsubsection {\arabic{subsubsection}}
		
		\makeatletter
		\renewcommand{\thefigure}{S\@arabic\c@figure}
		\renewcommand{\theequation}{S\@arabic\c@equation}
		\makeatother
		
		\setcounter{section}{0}    
		\setcounter{figure}{0}    
		\setcounter{equation}{0}
		\renewcommand\theequation{S\arabic{equation}}		
		
		
		
		
		

	\section {The existence analysis of the coexistent equilibrium}\label{coexistence equilibrium}
	For the higher-order GLV model~(\ref{eq:GLV_general}) with $S$ species, governed by the dynamical equation:
	\begin{equation}
	\frac{d{N}_i}{dt} = N_i \left( r_i + \sum_{j=1}^S \alpha_{ij}N_j + \sum_{j,k=1}^S \beta_{ijk}N_jN_k \right) + d_i, \quad i, j, k=1,2,\dots,S.
		\label{eq:GLV_general}
	\end{equation}
	where $\alpha_{ij}<0,~\beta_{ijk}<0$ and $\alpha_{ij}>0,~\beta_{ijk}>0$ represent the competitive and cooperative interactions among the species, or the system~(\ref{eq:GLV_general}) coexists a mixture of both interaction types.
	
	At steady state, from $\dot{N}_i =0$, we can derive the following algebraic equation for the equilibrium $N^*$:
	\begin{equation}
		N_i^* \left( r_i + \sum_{j=1}^S \alpha_{ij}N_j^* + \sum_{j,k=1}^S \beta_{ijk}N_j^*N_k^* \right) + d_i=0.
		\label{eq:equilibrium_condition}
	\end{equation}
	The trivial equilibrium $N_i^*=0$ exists when $d_i=0$, but it is ecologically irrelevant. Hence, we only consider the coexistent equilibrium $N_i^*$.  
	
	To analyze the existence of coexistence fixed points for the higher-order GLV model~(\ref{eq:GLV_general}) using Brouwer's Fixed-Point Theorem. Let's consider the simplest case and take three species as examples. Next, we will prove the existence of a coexistence fixed point, $N_i^*$, such that $N_i^* > 0$ for all $i$. Let $D \subset \mathbb{R}^3$ be a non-empty compact convex set, and $f: D \to D$ a continuous function. Then $f$ has at least one fixed point in $D$. Then, we need: (1) Construct a bounded convex region $D \subset \mathbb{R}^3_+$ where trajectories remain bounded; (2) Define a continuous map $f: D \to D$ whose fixed points correspond to equilibria; (3) Apply Brouwer’s theorem to guarantee existence. Here, we need to construct the domain $D$. Firstly, to guarantee boundedness of trajectories, we assume the system~(\ref{eq:GLV_general}) is dissipative. Whereas negative higher-order self-interactions ($\beta_{iii} < 0$) dominate at high population abunance and $d_i$ are finite for the system~(\ref{eq:GLV_general}). Then, there exists $\delta > 0$ such that for all $N_i \geq \delta$, $\quad r_i + \sum_{j} \alpha_{ij}N_j + \sum_{j,k} \beta_{ijk}N_jN_k < 0$. Secondly, let $D = [\epsilon, \delta]^3 \subset \mathbb{R}^3_+$, where $\epsilon > 0$ which is lower bound to exclude extinction, and $\delta > 0$ which is upper bound from dissipativity. This requires that the vector field points inward on $\partial D$: if $N_i = \epsilon$, then $\dot{N}_i \geq 0$; On the contrary, if $N_i = \delta$, then $\dot{N}_i \leq 0$. Next, we also need define the \textbf{time-1 map} $f: D \to D$ as $f(\mathbf{N}(t)) =\mathbf{N}(t+1)$, which advances the system by one time unit. By dissipativity and inward-pointing on $\partial D$, $f$ maps $D$ to itself, and continuity follows from the system~(\ref{eq:GLV_general}) smoothness. According to Brouwer’s theorem, $f$ has at least one fixed point $\mathbf{N}^* \in D$, satisfying $\dot{\mathbf{N}} = 0$, corresponding to an equilibrium $N_i^*$ in $D$. To ensure coexistence of equilibrium $N_i^* > 0$, the conditions 
	$d_i > 0$ and $\epsilon$ is sufficiently small are satisfied, then fixed points can lie on $N_i^* \subset \mathbb{R}^3_+$.
	
	\section {The local stability analysis of the coexistent equilibrium}\label{Jacobian}
	In order to determine the local asymptotic stability of system~(\ref{eq:GLV_general}), we need to linearise the system~(\ref{eq:GLV_general}) at the equilibrium point $N^*$. Define perturbation variables $x_i=N_i - N_i^*$, leading to the linear approximation:
	\begin{equation}
		\dot{x}_i = Jx.
		\label{eq:Jacobian}
	\end{equation}
	where the Jacobian matrix $J\in\mathbb{R}^{S\times S}$ has elements,
	\begin{equation}
		J_{lm} = \frac{\partial \dot{N}_l}{\partial N_m} \bigg|_{\mathbf{N}=\mathbf{N}^*}  = \delta_{lm} \left[ r_l + \sum_{j=1}^S \alpha_{lj}N_j^* + \sum_{j,k=1}^S \beta_{ljk}N_j^*N_k^* \right] + N_l^* \left( \alpha_{lm} + 2\sum_{k=1}^S \beta_{lmk}N_k^* \right).
		\label{eq:Jacobian_raw}
	\end{equation}
	Substitute \eqref{eq:equilibrium_condition} into \eqref{eq:Jacobian_raw} to obtain:
	\begin{equation}
		J_{lm} = \delta_{lm} \left( -\frac{d_l}{N_l^*} \right) + N_l^* \left( \alpha_{lm} + 2\sum_{k=1}^S \beta_{lmk}N_k^* \right).
		\label{eq:Jacobian_simplified}
	\end{equation}
	Due to local stability of the equilibrium $N_i^*$ is determined by the eigenvalues of Jacobian matrix $J$. Hence, if all the real parts of the eigenvalues of Jacobian matrix $J$ are less than 0, then system~(\ref{eq:GLV_general}) is locally asymptotically stable near the equilibrium point $N_i^*$. Otherwise, it is unstable.
	
	\section {The global stability analysis of the coexistent equilibrium}\label{global stability}
	To determine the global stability of system~(\ref{eq:GLV_general}), we construct the relative entropy function as a Lyapunov function :
	\begin{equation}
		V(\mathbf{N}) = \sum_{i=1}^n \left( N_i - N_i^* - N_i^* \ln \frac{N_i}{N_i^*} \right).
		\label{eq:Lyapunov_function}
	\end{equation}
	
	Differentiate $V$ with respect to time and substitute \eqref{eq:GLV_general}, we get
	\begin{equation}
		\frac{dV}{dt} = \sum_{i=1}^S \left(1 - \frac{N_i^*}{N_i} \right) \dot{N}_i = \sum_{i=1}^S \left( 1 - \frac{N_i^*}{N_i} \right) \left[ N_i \left( r_i + \sum_{j=1}^S \alpha_{ij}N_j + \sum_{j,k=1}^S \beta_{ijk}N_jN_k \right) + d_i \right].
		\label{eq:Lyapunov_function1}
	\end{equation}
	
	Substituting the equation \ref{eq:equilibrium_condition} into \ref{eq:Lyapunov_function1}, which is satisfied by the equilibrium point, we obtain
	\begin{equation}
		\frac{dV}{dt} = \sum_{i,j=1}^S \alpha_{ij} (N_i - N_i^*)(N_j - N_j^*) + \sum_{i,j,k=1}^S \beta_{ijk} \left[ N_i^* (N_j - N_j^*)(N_k - N_k^*) + (N_i - N_i^*)(N_j - N_j^*)(N_k - N_k^*) \right].
		\label{eq:dVdt_expanded}
	\end{equation}
	If the condition $\frac{dV}{dt} \leqslant 0$ is satisfied, then the coexistent equilibrium $N_i^*$ of system~(\ref{eq:GLV_general}) is globally stable.
	
	\section{Simulation details of the main text figures}\label{Simulation details}
	In Fig.~\ref{pairwise higher-order}: $r_{i}=0.3$, $d_{i}=0.001$, $(i, j, k= 1,\cdots,32)$. In Fig.~\ref{pairwise higher-order}b: $\alpha_{ij}=\mathcal{N}(\mu_{1}, \sigma_{1})$, $\beta_{ijk}=0$. In Fig.~\ref{pairwise higher-order}c: $\alpha_{ij}=\mathcal{N}(\mu_{1}, \sigma_{1})$, $\beta_{ijk}=\mathcal{N}(-0.3, 0.3)$. In Fig.~\ref{pairwise higher-order}d: $\alpha_{ij}=\mathcal{N}(-0.2, 0.2)$, $\beta_{ijk}=0$. In Fig.~\ref{pairwise higher-order}e: $\alpha_{ij}=\mathcal{N}(-0.6, 0.8)$, $\beta_{ijk}=0$. In Fig.~\ref{pairwise higher-order}f: $\alpha_{ij}=\mathcal{N}(-0.2, 0.2)$, $\beta_{ijk}=\mathcal{N}(-0.3, 0.3)$. In Fig.~\ref{pairwise higher-order}g: $\alpha_{ij}=\mathcal{N}(-0.6, 0.8)$, $\beta_{ijk}=\mathcal{N}(-0.3, 0.3)$.
	
	In Fig.~\ref{Bifurcation-mu2}: $r_{i}=0.8$, $d_{i}=0.001$, $\alpha_{ij}=\mathcal{N}(-0.1, 0.1)$, $\beta_{ijk}=\mathcal{N}(\mu_{2}, 0.4)$, $(i, j, k=1,\cdots,32)$. In Fig.~\ref{Bifurcation-mu2}b-d: $\mu_{2}=-0.45$.
	
	Model settings in Fig.~\ref{FigRank}a (Only pairwise): For bee community, $r_{i}=0.3$, $d_{i}=0.001$, $\alpha_{ij}=\mathcal{N}(-0.3, 0.2)$, $\beta_{ijk}=0$, $(i, j, k= 1,\cdots,32)$; For grasshopper community, $r_{i}=0.3$, $d_{i}=0.001$, $\alpha_{ij}=\mathcal{N}(-0.3, 0.19)$, $\beta_{ijk}=0$, $(i, j, k= 1,\cdots,28)$; For bird community, $r_{i}=0.45$, $d_{i}=0.001$, $\alpha_{ij}=\mathcal{N}(-0.3, 0.15)$, $\beta_{ijk}=0$, $(i, j, k= 1,\cdots,23)$; For bat community, $r_{i}=0.4$, $d_{i}=0.001$, $\alpha_{ij}=\mathcal{N}(-0.38, 0.25)$, $\beta_{ijk}=0$, $(i, j, k= 1,\cdots,21)$; In the K-S test, the probabilities (pvalues) that the simulation results involving only pairwise interactions and the corresponding observed data come from the different distributions are:
	$p_\text{bee}=1.21\times10^{-4}$, $p_\text{grasshopper}=9.16\times10^{-12}$, $p_\text{bird}=1.62\times10^{-9}$, $p_\text{bat}=5.53\times10^{-7}$. 
	Model settings in Fig.~\ref{FigRank}a (Pairwise \& higher-order): For bee community, $r_{i}=0.3$, $d_{i}=0.001$, $\alpha_{ij}=\mathcal{N}(-0.3, 0.2)$, $\beta_{ijk}=\mathcal{N}(-0.15, 0.3)$, $(i, j, k= 1,\cdots,32)$; For grasshopper community, $r_{i}=0.3$, $d_{i}=0.001$, $\alpha_{ij}=\mathcal{N}(-0.3, 0.19)$, $\beta_{ijk}=\mathcal{N}(-0.3, 0.45)$, $(i, j, k= 1,\cdots,28)$; For bird community, $r_{i}=0.45$, $d_{i}=0.001$, $\alpha_{ij}=\mathcal{N}(-0.3, 0.15)$, $\beta_{ijk}=\mathcal{N}(-0.25, 0.4)$, $(i, j, k= 1,\cdots,23)$; For bat community, $r_{i}=0.4$, $d_{i}=0.001$, $\alpha_{ij}=\mathcal{N}(-0.38, 0.25)$, $\beta_{ijk}=\mathcal{N}(-0.33, 0.3)$, $(i, j, k= 1,\cdots,21)$. In the K-S test, the probabilities (pvalues) that the simulation results involving pairwise \& higher-order interactions and the corresponding observed data come from the same distributions are:
	$p_\text{bee}=0.38$, $p_\text{grasshopper}=0.49$, $p_\text{bird}=0.59$, $p_\text{bat}=0.30$. The Shannon entropies of the observed data and simulation results for each ecological community are: $H_{\text{Obs(ODEs)}}^{\text{bee}} =3.66(3.45)$, $H_{\text{Obs(ODEs)}}^{\text{grasshopper}} =3.45(3.22)$, $H_{\text{Obs(ODEs)}}^{\text{bird}} =3.43(2.92)$, $H_{\text{Obs(ODEs)}}^{\text{bat}} =3.26(3.03)$. 
	Model settings in Fig.~\ref{FigRank}b (plankton): $r_{i}=0.55$, $d_{i}=0.001$, $\alpha_{ij}=\mathcal{N}(-0.1, 0.1)$, $\beta_{ijk}=\mathcal{N}(-0.4, 0.4)$, $(i, j, k= 1,\cdots,39)$. 
	Model settings in Fig.~\ref{FigRank}b (bird): $r_{i}=0.2$, $d_{i}=0.001$, $\alpha_{ij}=\mathcal{N}(-0.1, 0.1)$, $\beta_{ijk}=\mathcal{N}(-0.2, 0.18)$, $(i, j, k= 1,\cdots,16)$. 
	Model settings in Fig.~\ref{FigRank}b (aquatic invertebrate): $r_{i}=0.35$, $d_{i}=0.001$, $\alpha_{ij}=\mathcal{N}(-0.1, 0.1)$, $\beta_{ijk}=\mathcal{N}(-0.2, 0.15)$, $(i, j, k= 1,\cdots,10)$. 
	Model settings in Fig.~\ref{FigRank}b (mosquito): $r_{i}=0.2$, $d_{i}=0.001$, $\alpha_{ij}=\mathcal{N}(-0.1, 0.1)$, $\beta_{ijk}=\mathcal{N}(-0.25, 0.23)$, $(i, j, k= 1,\cdots,5)$. 
	In the K-S test, the probabilities (pvalues) that the simulation results involving pairwise \& higher-order interactions and the corresponding observed data come from the same distributions are:
	$p_\text{plankton}=0.71$, $p_\text{bird}=0.91$, $p_\text{aquatic invertebrate}=1.00$, $p_\text{mosquito}=1.00$. The Shannon entropies of the observed data and simulation results for each ecological community are: $H_{\text{Obs(ODEs)}}^{\text{plankton}} =4.66(4.64)$, $H_{\text{Obs(ODEs)}}^{\text{bird}} =3.04(3.08)$, $H_{\text{Obs(ODEs)}}^{\text{aquatic invertebrate}} =1.67(1.81)$, $H_{\text{Obs(ODEs)}}^{\text{mosquito}} =1.36(1.42)$.
	Model settings in Fig.~\ref{FigRank}c: $d_{i}=0.001$, $\alpha_{ij}=\mathcal{N}(-0.1, 0.1)$. 
	Model settings in Fig.~\ref{FigRank}c ($S=6$): $r_{i}=0.3$, $\beta_{ijk}=\mathcal{N}(-0.2, 0.15)$, $(i, j, k= 1,\cdots,6)$. 
	Model settings in Fig.~\ref{FigRank}c ($S=7$): $r_{i}=0.3$, $\beta_{ijk}=\mathcal{N}(-0.2, 0.16)$, $(i, j, k= 1,\cdots,7)$. 
	Model settings in Fig.~\ref{FigRank}c ($S=8$): $r_{i}=0.3$, $\beta_{ijk}=\mathcal{N}(-0.25, 0.15)$, $(i, j, k= 1,\cdots,8)$. 
	Model settings in Fig.~\ref{FigRank}c ($S=9$): $r_{i}=0.18$, $\beta_{ijk}=\mathcal{N}(-0.2, 0.15)$, $(i, j, k= 1,\cdots,9)$. 
	Model settings in Fig.~\ref{FigRank}c ($S=11$): $r_{i}=0.2$, $\beta_{ijk}=\mathcal{N}(-0.2, 0.15)$, $(i, j, k= 1,\cdots,11)$. 
	Model settings in Fig.~\ref{FigRank}c ($S=15$): $r_{i}=0.18$, $\beta_{ijk}=\mathcal{N}(-0.2, 0.15)$, $(i, j, k= 1,\cdots,15)$. 
	Model settings in Fig.~\ref{FigRank}c ($S=16$): $r_{i}=0.2$, $\beta_{ijk}=\mathcal{N}(-0.2, 0.18)$, $(i, j, k= 1,\cdots,16)$. 
	Model settings in Fig.~\ref{FigRank}c ($S=17$): $r_{i}=0.18$, $\beta_{ijk}=\mathcal{N}(-0.2, 0.15)$, $(i, j, k= 1,\cdots,17)$. 
	Model settings in Fig.~\ref{FigRank}c ($S=18$): $r_{i}=0.2$, $\beta_{ijk}=\mathcal{N}(-0.2, 0.22)$, $(i, j, k= 1,\cdots,18)$. 
	Model settings in Fig.~\ref{FigRank}c ($S=23$): $r_{i}=0.24$, $\beta_{ijk}=\mathcal{N}(-0.19, 0.32)$, $(i, j, k= 1,\cdots,23)$. 
	Model settings in Fig.~\ref{FigRank}c ($S=30$): $r_{i}=0.28$, $\beta_{ijk}=\mathcal{N}(-0.25, 0.3)$, $(i, j, k= 1,\cdots,30)$. 
	Model settings in Fig.~\ref{FigRank}c ($S=34$): $r_{i}=0.3$, $\beta_{ijk}=\mathcal{N}(-0.23, 0.4)$, $(i, j, k= 1,\cdots,34)$. 
	Model settings in Fig.~\ref{FigRank}c ($S=36$): $r_{i}=0.32$, $\beta_{ijk}=\mathcal{N}(-0.23, 0.4)$, $(i, j, k= 1,\cdots,36)$. 
	Model settings in Fig.~\ref{FigRank}c ($S=37$): $r_{i}=0.30$, $\beta_{ijk}=\mathcal{N}(-0.22, 0.39)$, $(i, j, k= 1,\cdots,37)$. 
	Model settings in Fig.~\ref{FigRank}c ($S=38$): $r_{i}=0.33$, $\beta_{ijk}=\mathcal{N}(-0.25, 0.3)$, $(i, j, k= 1,\cdots,38)$. 
	Model settings in Fig.~\ref{FigRank}c ($S=51$): $r_{i}=0.6$, $\beta_{ijk}=\mathcal{N}(-0.45, 0.55)$, $(i, j, k= 1,\cdots,51)$.

	\clearpage
	\section{Supplemental Figures}
	\begin{figure}[h]
		\centering
		\includegraphics[width=17cm]{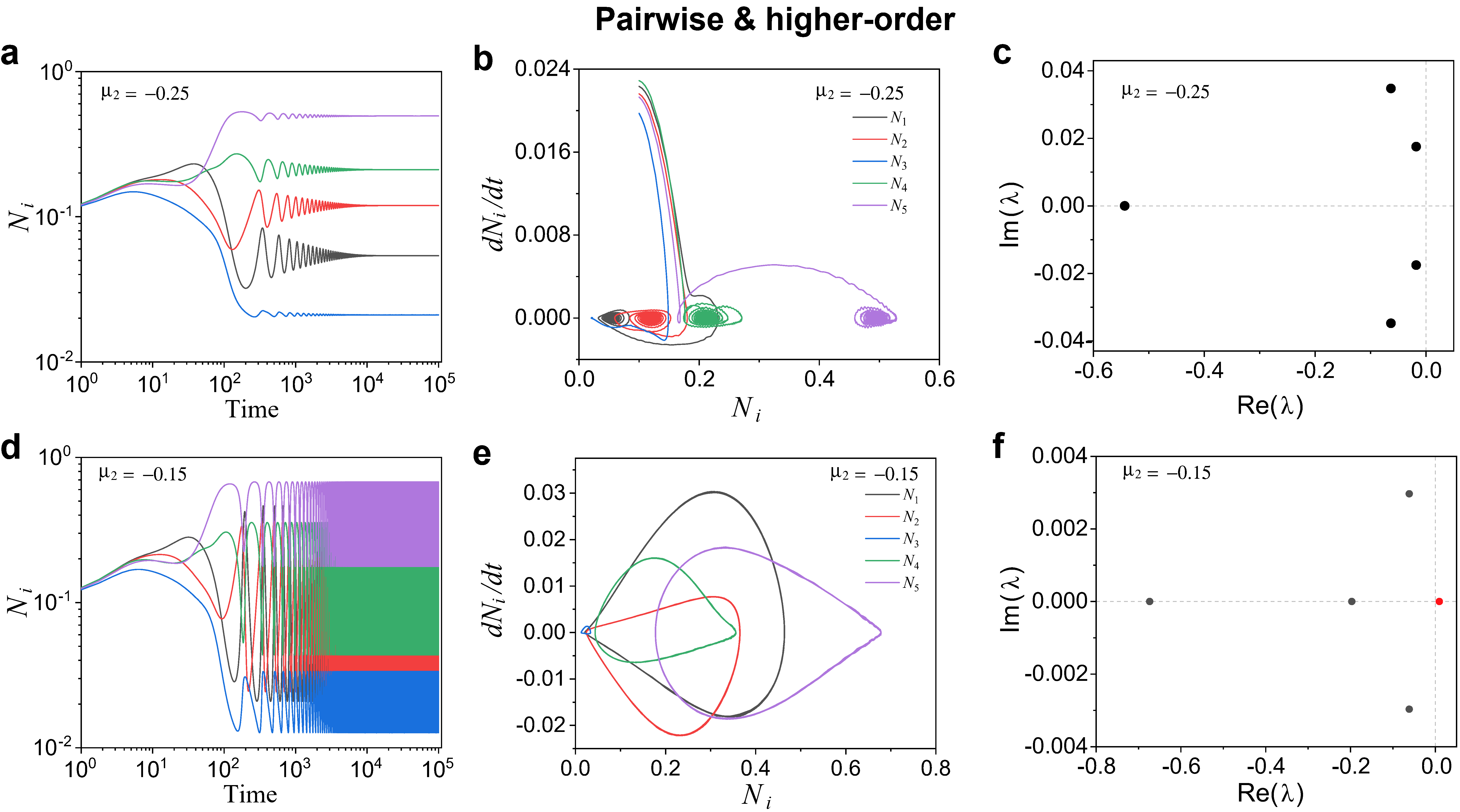}
		\caption{\label{stability} The stability of the coexistent equilibrium. (a-b, d-e) Time courses and phase diagrams in the scenario involving pairwise and higher-order interactions, respectively. 
			(c, f) The eigenvalues of the Jacobian of this community when the scenario involving pairwise and higher-order interactions is considered.    (c) The coexistent equilibrium is stable when all the eigenvalues have a negative real part corresponding to \ref{stability}(a-b). (f) The red dotted displays the eigenvalues with a positive real part, indicating instabilitycorresponding to \ref{stability}(d-e). The simulations involve $S = 5$ species. In (a-f): $r_{i}=0.4$, $d_{i}=0.001$, $\alpha_{ij}=\mathcal{N}(-0.25, 0.1)$ $\beta_{ijk}=\mathcal{N}(\mu_{2}, 0.3)$, $i, j , k= 1,\dots, 5$; In (a-c): $\mu_{2}=-0.25$; In (d-f): $\mu_{2}=-0.15$.}	
	\end{figure}

	\begin{figure}[h]
		\centering
		\includegraphics[width=15cm]{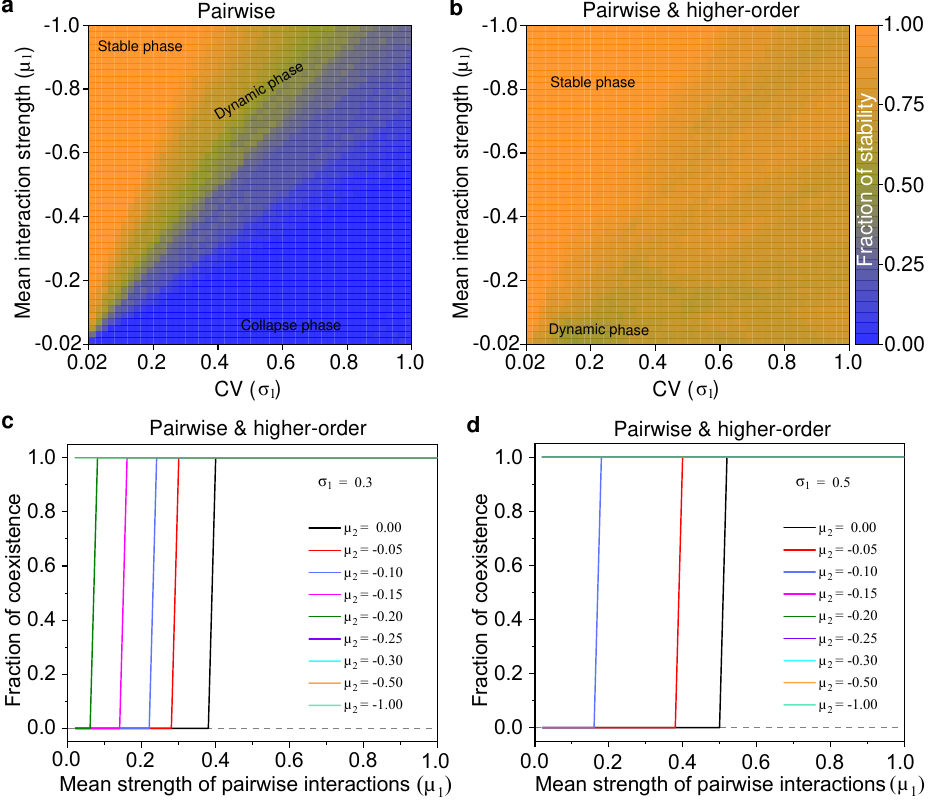}
		\caption{\label{randfig} The influence of stochasticity on species coexistence and stability. (a-b) Phase diagrams in the scenario involving pairwise and higher-order interactions, respectively. The species’ stability fraction in each pixel was calculated from 60 random repeats. 
		(c-d) The coexisting fraction of species is introduced by higher-order interactions. The fraction of species that coexist under different mean strengths of higher-order interactions. In (a-b) the parameter values are provided in Fig.~\ref{pairwise higher-order}. In (c-d): $r_{i}=0.3$, $d_{i}=0.001$, $\beta_{ijk}=\mathcal{N}(\mu_{2}, 0.3)$, $i, j , k= 1,\dots, 32$; In (c): $\alpha_{ij}=\mathcal{N}(\mu_{1}, 0.3)$; In (d): $\alpha_{ij}=\mathcal{N}(\mu_{1}, 0.5)$.}	
	\end{figure}
	
	\begin{figure}[h]
		\centering
		\includegraphics[width=17cm]{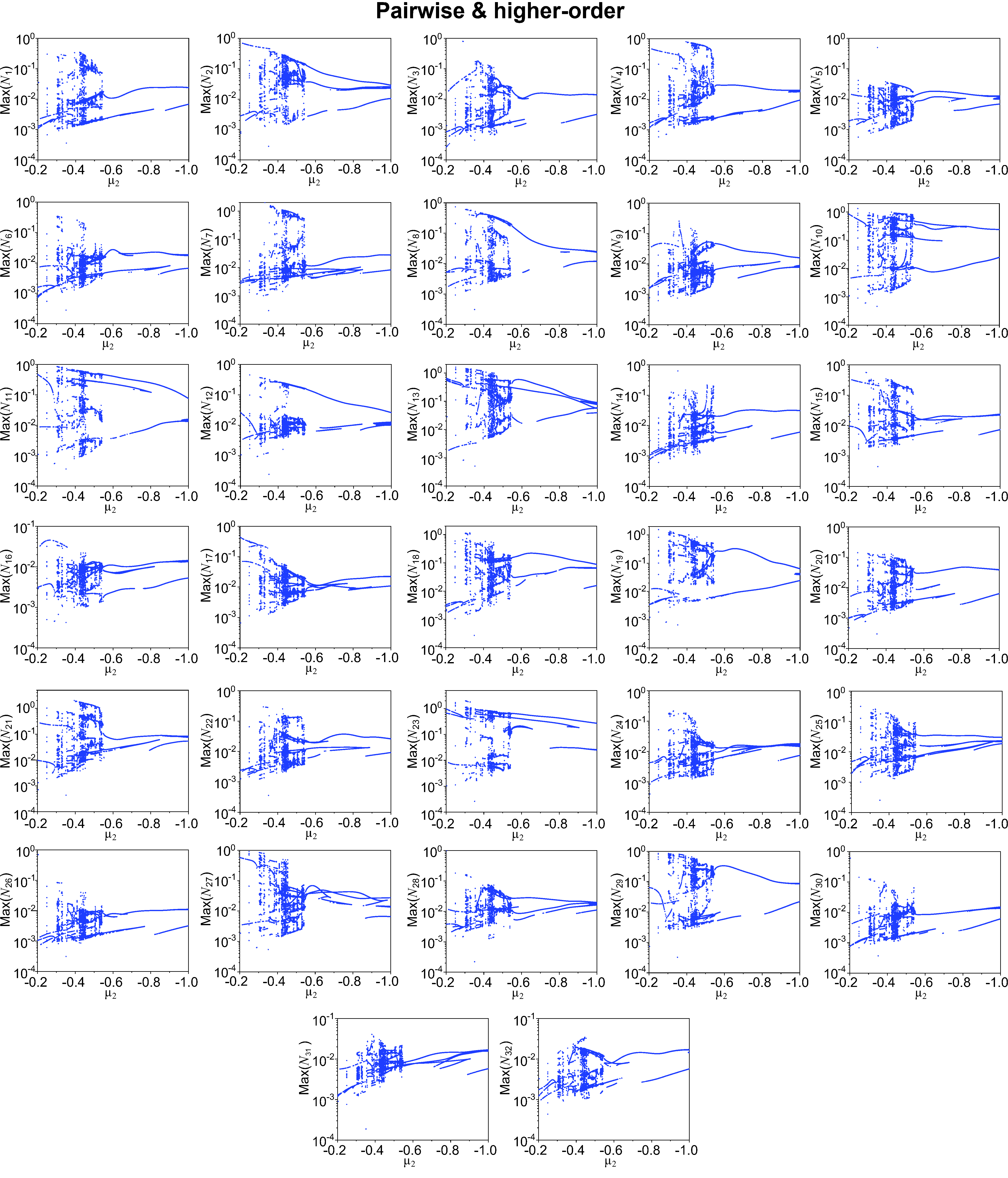}
		\caption{\label{SF-Bifurcation-mu2} Bifurcation diagrams of the system corresponding to the variation of mean interaction strength  $\mu_{2}$. The parameter values are provided in Fig.~\ref{Bifurcation-mu2}. }	
	\end{figure}
	
	\begin{figure}[h]
		\centering
		\includegraphics[width=17cm]{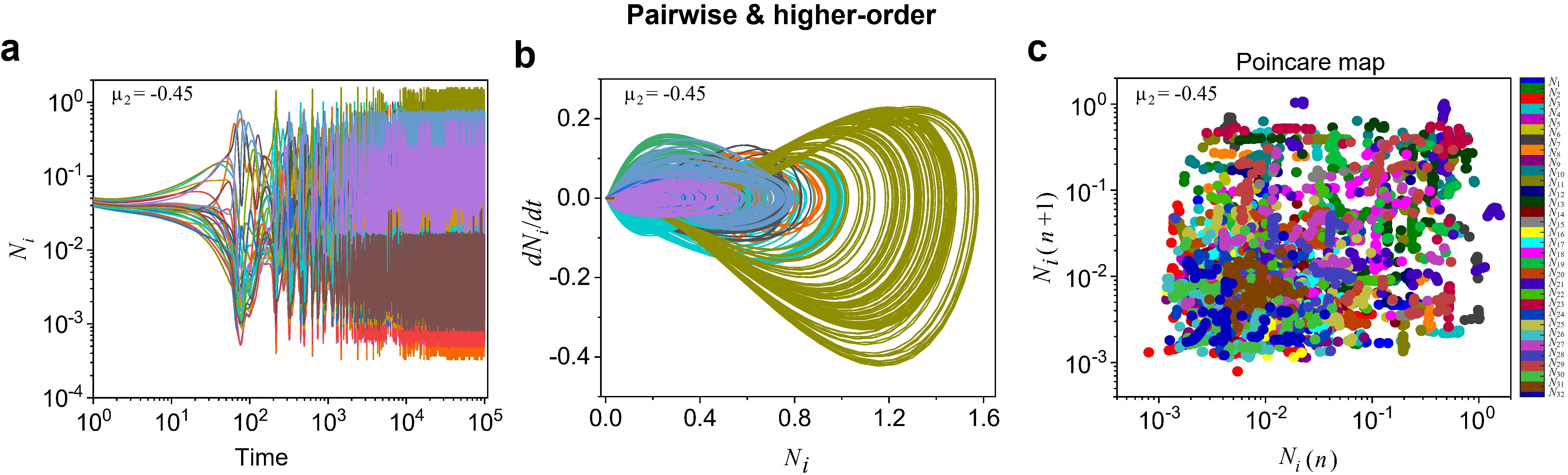}
		\caption{\label{SF-Bifurcation-mu2-phase} Complex dynamics are triggered by both pairwise and higher-order interactions. (a-c) Chaotic dynamics in time course, 2-D projections of phase space and Poincare map, corresponding to the dynamic variation in the bifurcation diagram of Figs.~\ref{Bifurcation-mu2}a,~\ref{SF-Bifurcation-mu2} at $\mu_{2}=-0.45$. The other parameter values are provided in Fig.~\ref{Bifurcation-mu2}. }	
	\end{figure}
	
	\begin{figure}[h]
		\centering
		\includegraphics[width=15cm]{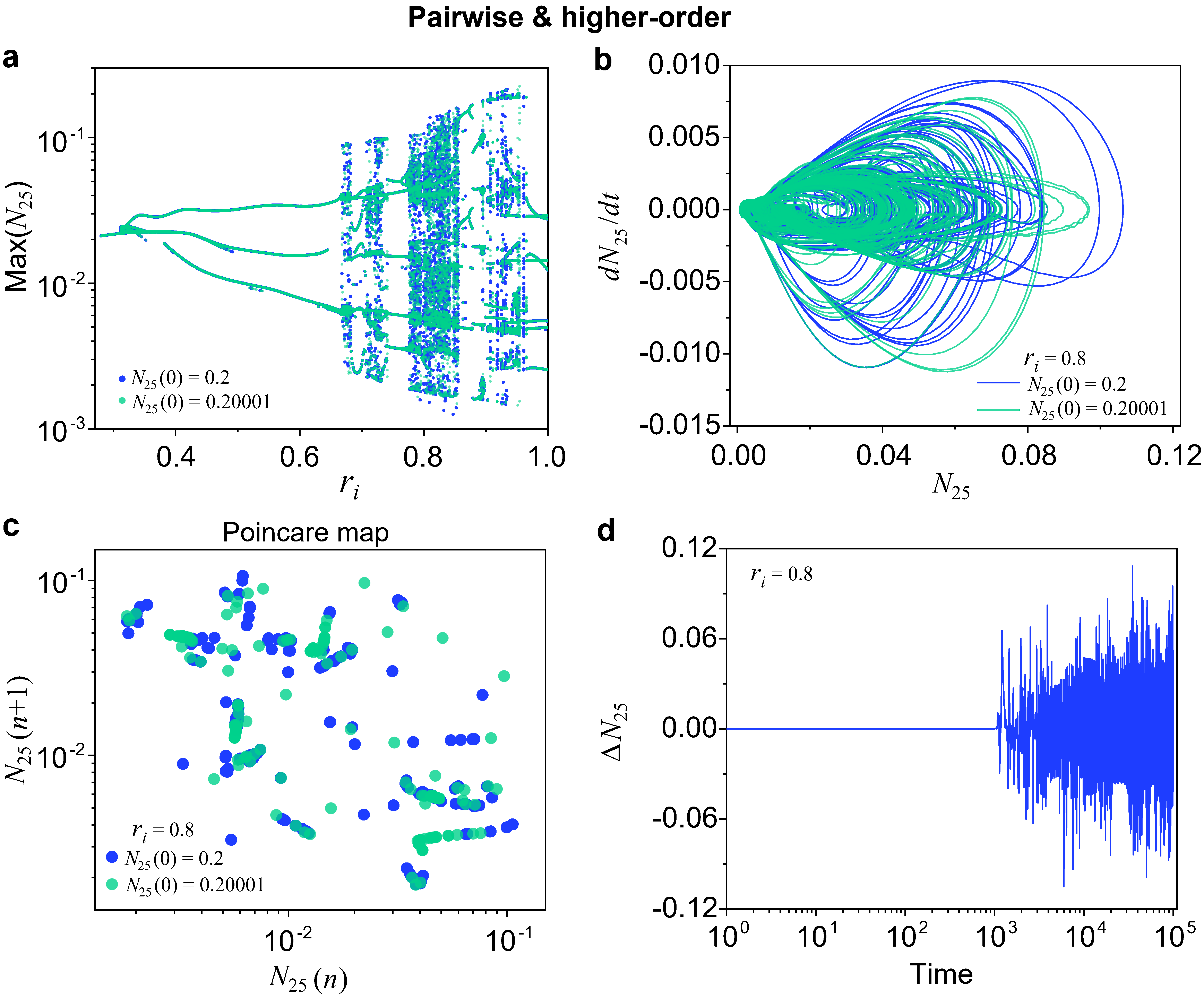}
		\caption{\label{bif_Poincare} The emergence of chaotic dynamics is induced by pairwise and higher-order interactions. (a-c) The blue and green dots represent the simulation results for initial conditions of $N_{25}(0)=0.2$ and $N_{25}(0)=0.20001$, respectively, with all other initial values set to $N_{i}(0)=0.2$ ($i=1,\cdots,32 $). (a) Bifurcation diagram of the system~(\ref{GLV}) corresponding to the variation of parameter $r$. (b-c) Chaotic dynamics in 2-D projections of phase space and Poincare map, corresponding to the dynamic variation in the bifurcation diagram of Fig.~\ref{bif_Poincare}a at $r=0.8$. (d) Sensitivity analysis on the dynamics of system~(\ref{GLV}) corresponding to the parametric conditions of Fig.~\ref{bif_Poincare}b-c. All simulations involve $S = 32$ species. In Fig.~\ref{bif_Poincare}: $d_{i}=0.001$, $\alpha_{ij}=\mathcal{N}(-0.1, 0.1)$, $\beta_{ijk}=\mathcal{N}(-0.45, 0.4)$, $(i, j, k=1,\cdots,32)$. In Fig.~\ref{bif_Poincare}b-d: $r_{i}=0.8$.}	
	\end{figure}
	
	\begin{figure}[h]
		\centering
		\includegraphics[width=17cm]{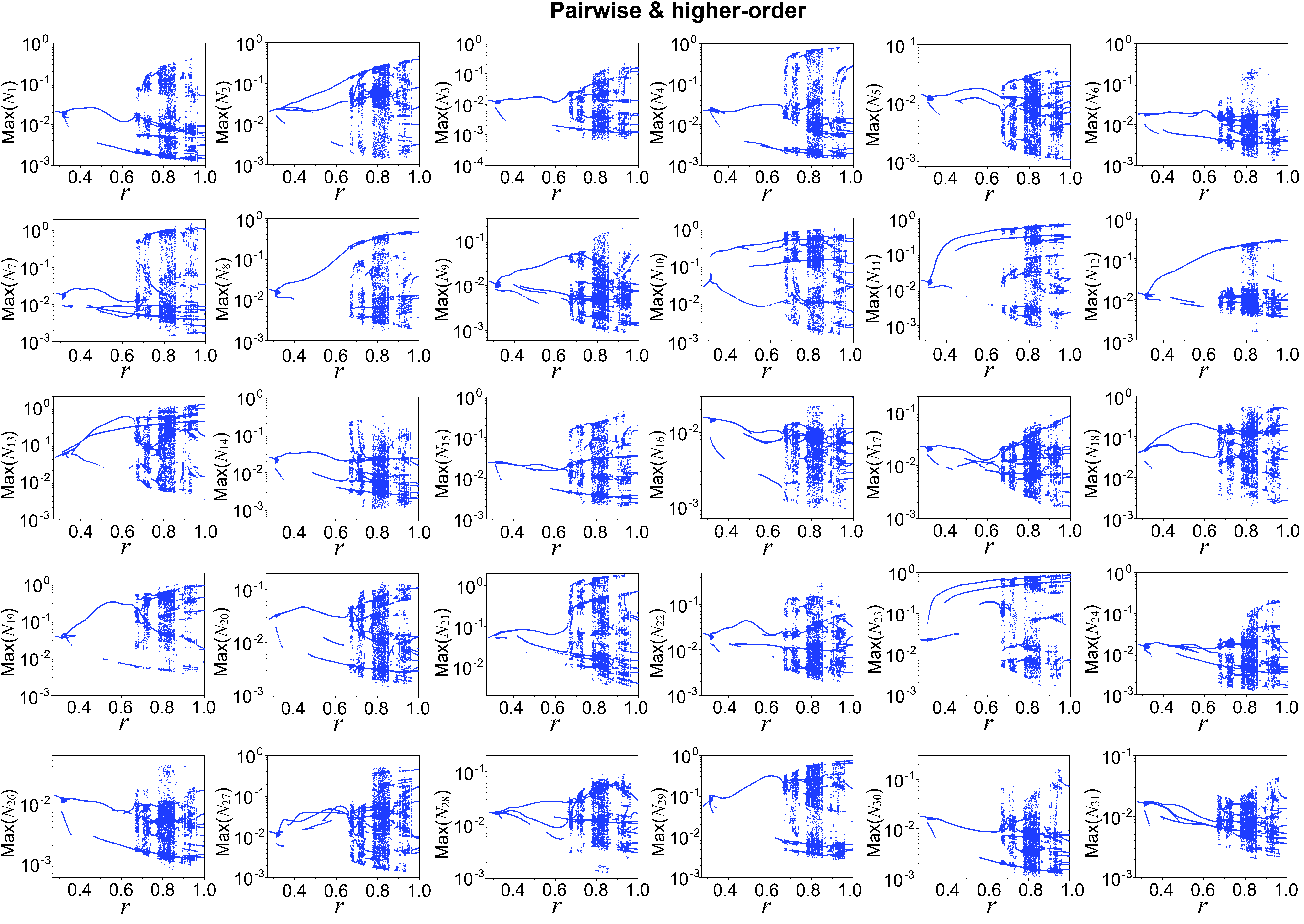}
		\caption{\label{bifurcation_SM00} Bifurcation diagrams of the system corresponding to the variation of parameter $r\equiv r_{i}$. The parameter values are provided in Fig.~\ref{bif_Poincare}. }	
	\end{figure}

	\begin{figure}[h]
		\centering
		\includegraphics[width=15cm]{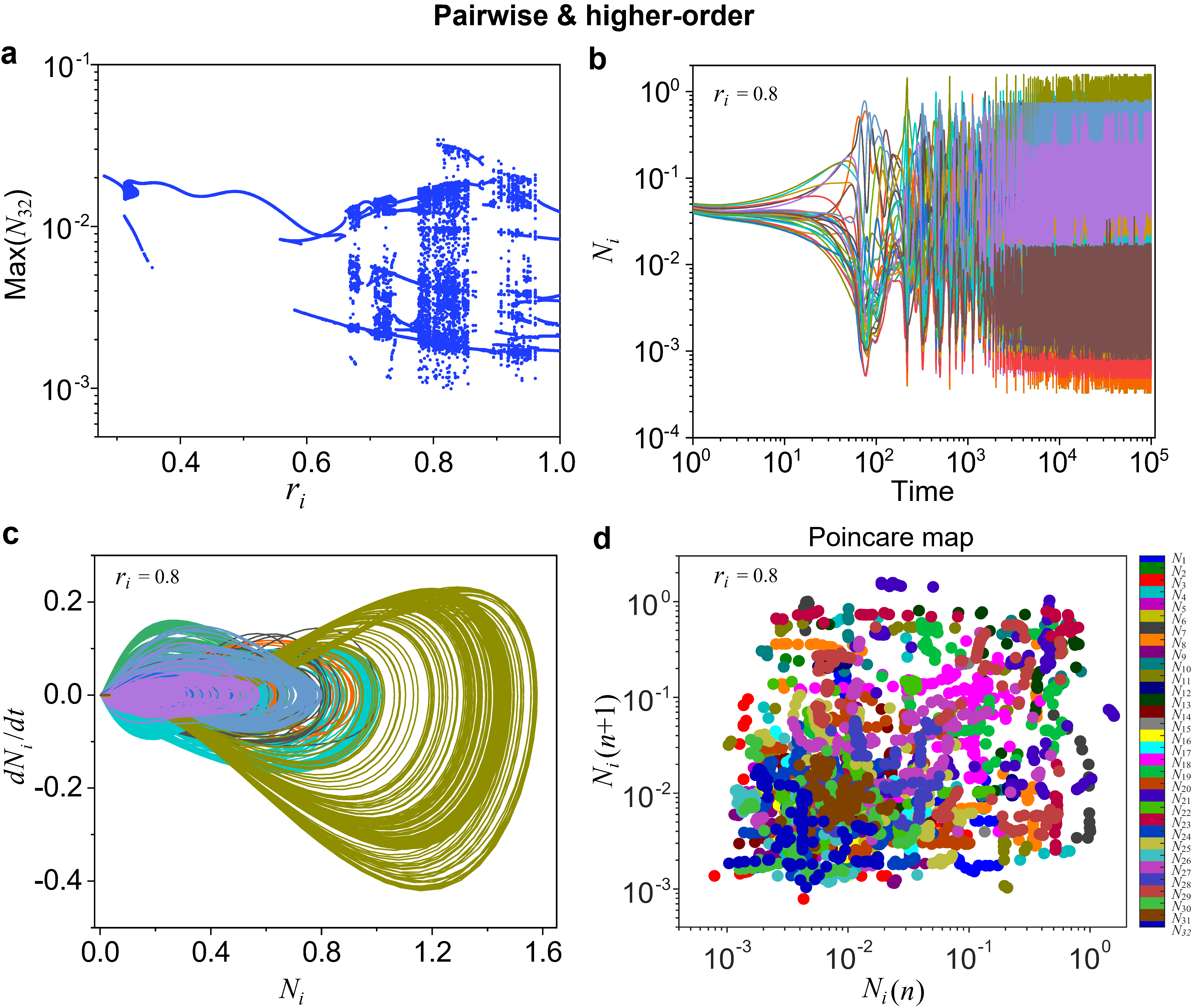}
		\caption{\label{bif_Poincare_SM} Complex dynamics are triggered by both pairwise and higher-order interactions. (a) Bifurcation diagram of the system corresponding to the variation of parameter $r$. (b-d) Chaotic dynamics in time course, 2-D projections of phase space and Poincare map, corresponding to the dynamic variation in the bifurcation diagram of Figs.~\ref{bif_Poincare}a,~\ref{bifurcation_SM00},~\ref{bif_Poincare_SM}a. The parameter values are provided in Fig.~\ref{bif_Poincare} at $r_{i}=0.8$. }	
	\end{figure}
	
	\begin{figure}[h]
		\centering
		\includegraphics[width=17cm]{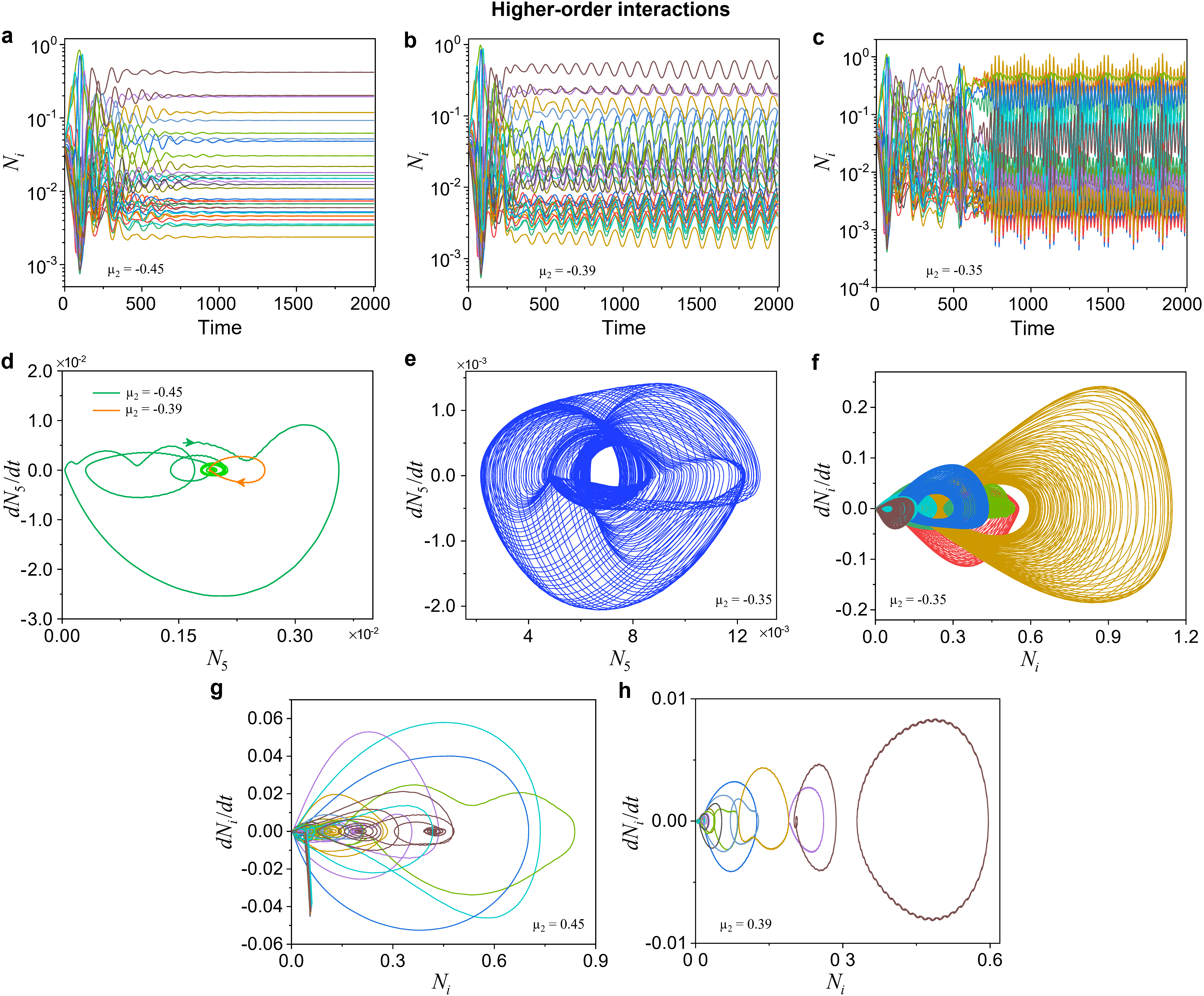}
		\caption{\label{self-organized dynamics} Self-organized biodiversity modes induced by only higher-order interactions. (a-c) Time courses of species abundances. (d-h) Representative trajectories of species' self-organized coexistence modes in 2-D projections of phase space. 
		All simulations involve $S = 32$ species. 
		In Fig.~\ref{self-organized dynamics}: $r_{i}=0.8$, $d_{i}=0.001$, $\alpha_{ij}=0$, $(i, j, k= 1,\cdots,32)$. In Fig.~\ref{self-organized dynamics}a, d(green): $\beta_{ijk}=\mathcal{N}(-0.45, 0.4)$. In Fig.~\ref{self-organized dynamics}b, d(orange): $\beta_{ijk}=\mathcal{N}(-0.39, 0.4)$. Fig.~\ref{self-organized dynamics}c, e, f: $\beta_{ijk}=\mathcal{N}(-0.35, 0.4)$.}	
	\end{figure}
	
	\begin{figure}[ht!]   
		\centering
		\includegraphics[width=16cm]{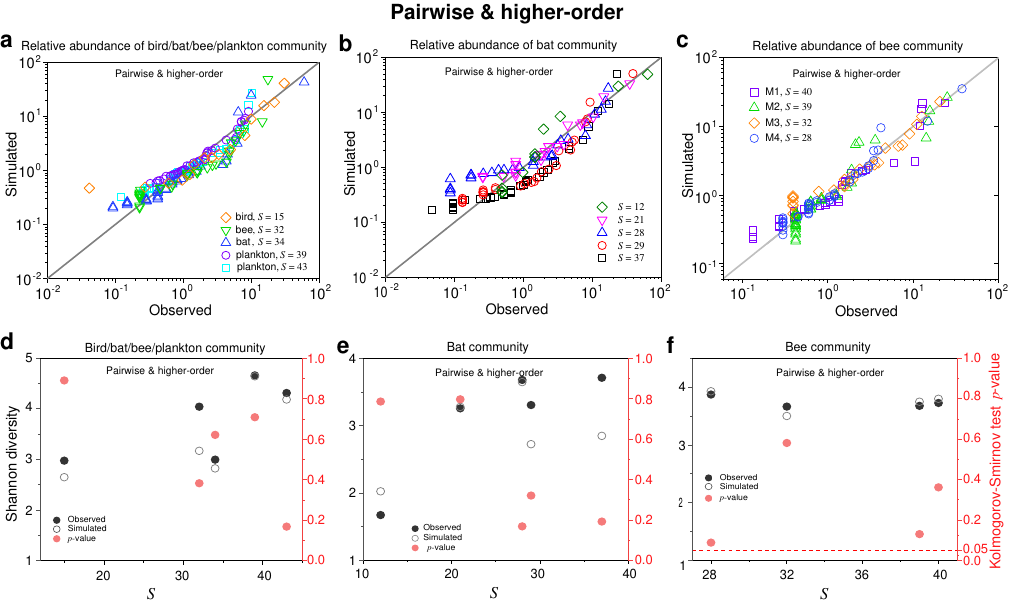}
		\caption{\label{elife-2020bee-2001batdata} Higher-order interaction illustrates the distribution pattern of the species’ in different ecological communities. 
			(a-c) A visual comparison of species distribution across the bird, bee, bat, plankton communities, and these observed data reported in existing studies~\cite{HubbellSP2001,FuhrmanJA2008,ClarkeFM2005,EstradaA2001,MichaelRoswell2021}, the simulated results were constructed from timestamp $ \textit{t} = 1.0\times10^{5} $ in the time series. 
			(d-f) The Shannon diversity and K-S test (\emph{p}-values) indicating whether the simulation results and the corresponding observed data come from identical distributions in (a-c), respectively. In the K-S test, using a significance threshold of 0.05, none of the \emph{p}-values indicate a statistically significant difference. In Fig.~\ref{elife-2020bee-2001batdata}a-c: $d_{i}=0.001$, $\alpha_{ij}=\mathcal{N}(-0.1, 0.1)$.
			In Fig.~\ref{elife-2020bee-2001batdata}a (brid): $r_{i}=0.25$, $\beta_{ijk}=\mathcal{N}(-0.22, 0.2)$, $(i, j, k= 1,\cdots,15)$. In Fig.~\ref{elife-2020bee-2001batdata}a (bee): $r_{i}=0.3$, $\beta_{ijk}=\mathcal{N}(-0.3, 0.34)$, $(i, j, k= 1,\cdots,32)$. In Fig.~\ref{elife-2020bee-2001batdata}a (bat): $r_{i}=0.25$, $\beta_{ijk}=\mathcal{N}(-0.3, 0.4)$, $(i, j, k= 1,\cdots,34)$. In Fig.~\ref{elife-2020bee-2001batdata}a (plankton, $S=39$): $r_{i}=0.55$, $\beta_{ijk}=\mathcal{N}(-0.4, 0.4)$, $(i, j, k= 1,\cdots,39)$. In Fig.~\ref{elife-2020bee-2001batdata}a (plankton, $S=43$): $r_{i}=0.55$, $\beta_{ijk}=\mathcal{N}(-0.39, 0.5)$, $(i, j, k= 1,\cdots,43)$. In Fig.~\ref{elife-2020bee-2001batdata}b (bat, $S=12$): $r_{i}=0.4$, $\beta_{ijk}=\mathcal{N}(-0.33, 0.27)$, $(i, j, k= 1,\cdots,12)$. In Fig.~\ref{elife-2020bee-2001batdata}b (bat, $S=21$): $r_{i}=0.4$, $\beta_{ijk}=\mathcal{N}(-0.33, 0.3)$, $(i, j, k= 1,\cdots,21)$. In Fig.~\ref{elife-2020bee-2001batdata}b (bat, $S=28$): $r_{i}=0.22$, $\beta_{ijk}=\mathcal{N}(-0.2, 0.43)$, $(i, j, k= 1,\cdots,28)$. In Fig.~\ref{elife-2020bee-2001batdata}b (bat, $S=29$): $r_{i}=0.26$, $\beta_{ijk}=\mathcal{N}(-0.32, 0.45)$, $(i, j, k= 1,\cdots,29)$. In Fig.~\ref{elife-2020bee-2001batdata}b (bat, $S=37$): $r_{i}=0.4$, $\beta_{ijk}=\mathcal{N}(-0.33, 0.35)$, $(i, j, k= 1,\cdots,37)$. In Fig.~\ref{elife-2020bee-2001batdata}c (bee, $S=40$): $r_{i}=0.2$, $\beta_{ijk}=\mathcal{N}(-0.15, 0.25)$, $(i, j, k= 1,\cdots,40)$. In Fig.~\ref{elife-2020bee-2001batdata}c (bee, $S=39$): $r_{i}=0.4$, $\beta_{ijk}=\mathcal{N}(-0.25, 0.4)$, $(i, j, k= 1,\cdots,39)$. In Fig.~\ref{elife-2020bee-2001batdata}c (bee, $S=32$): $r_{i}=0.4$, $\beta_{ijk}=\mathcal{N}(-0.25, 0.3)$, $(i, j, k= 1,\cdots,32)$. In Fig.~\ref{elife-2020bee-2001batdata}c (bee, $S=28$): $r_{i}=0.3$, $\beta_{ijk}=\mathcal{N}(-0.25, 0.3)$, $(i, j, k= 1,\cdots,28)$.}   
	\end{figure}

	\begin{figure}[ht!]   
		\centering
		\includegraphics[width=16cm]{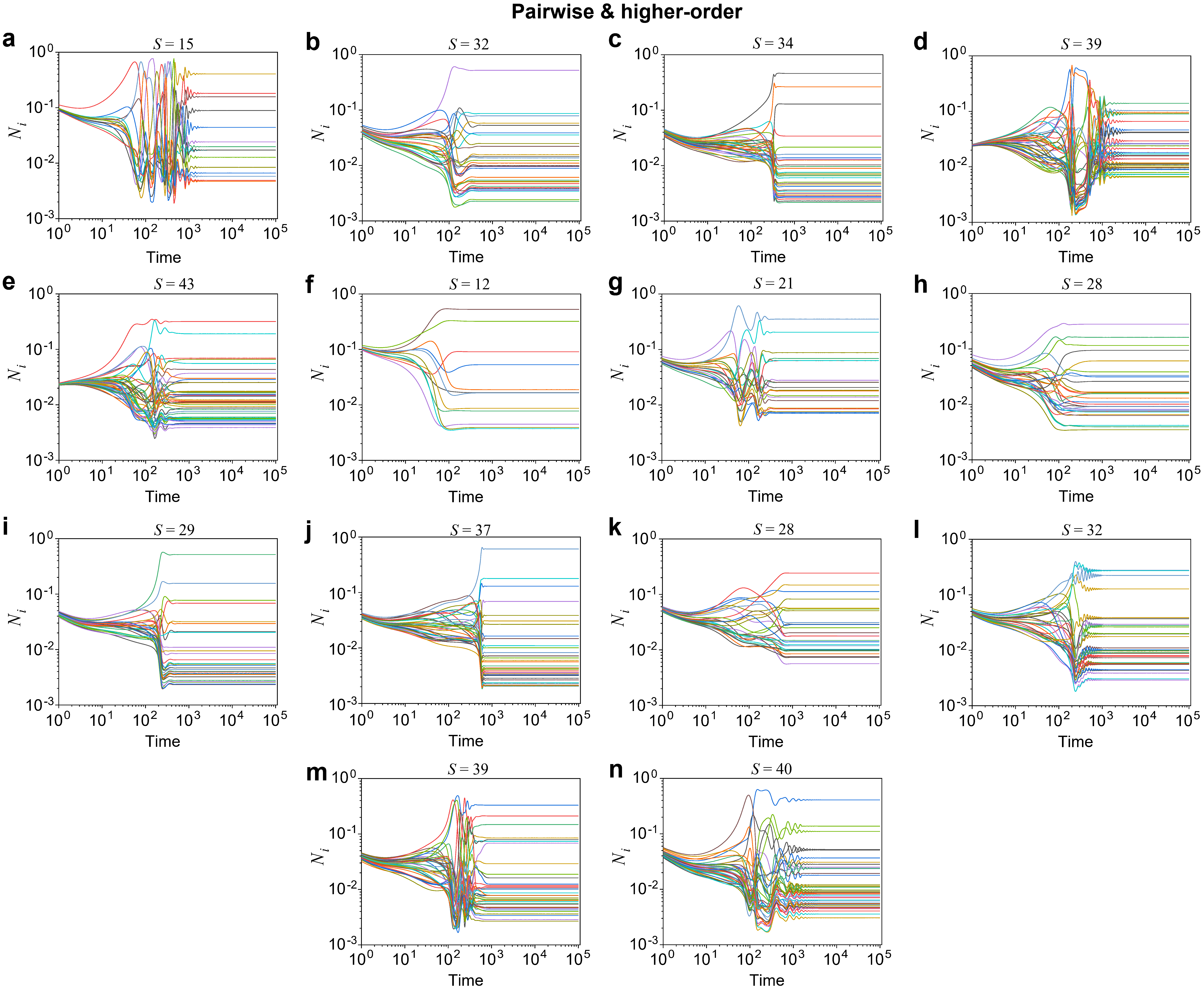}
		\caption{\label{Rank-abundance elife-2020bee-2001bat-timeseries} Higher-order interaction enables a wide range of consumer species to coexist. (a-n) Time courses of the species abundances simulated with system (1). The time series in (a-e), (f-j) and (k-n) correspond to that shown in Fig.~\ref{elife-2020bee-2001batdata}a-c, respectively. }   
	\end{figure}

	\begin{figure}[ht!]   
		\centering
		\includegraphics[width=16cm]{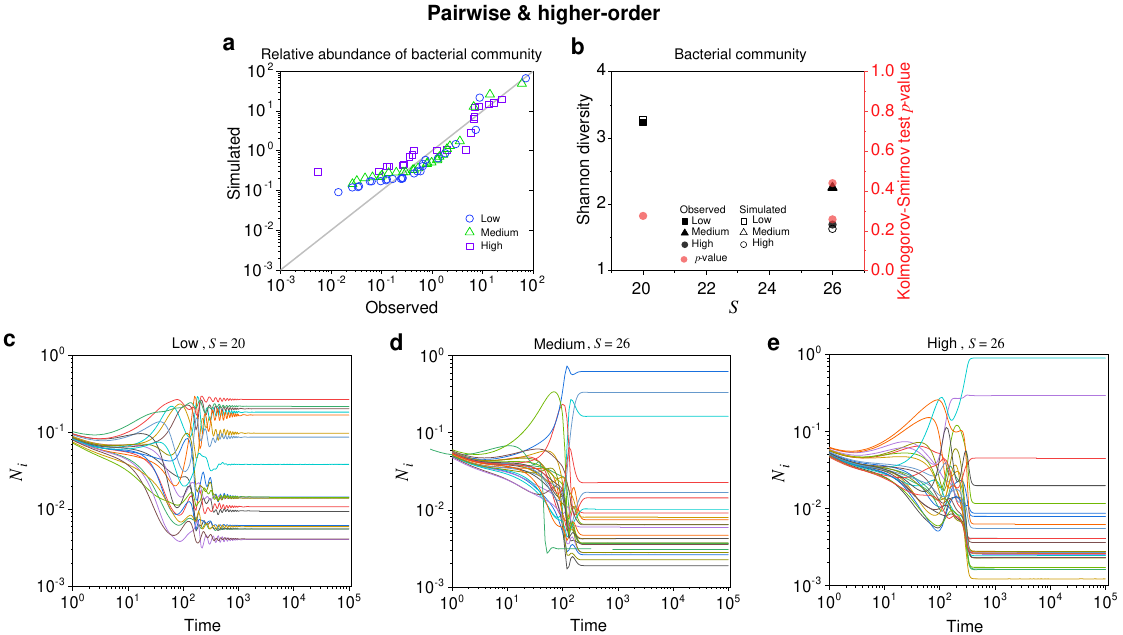}
		\caption{\label{Hudata} Higher-order interaction illustrates the distribution pattern of the species’ in bacterial community. (a) A visual comparison of species distribution across the bacterial community, and the observed data reported in existing studies~\cite{Hu2022}, the simulated results were constructed from timestamp $ \textit{t} = 1.0\times10^{5} $ in the time series. 
			(b) The Shannon diversity and K-S test (\emph{p}-values) indicating whether the simulation results and the corresponding observed data come from identical distributions in (a), and in the K-S test, using a significance threshold of 0.05, none of the \emph{p}-values indicate a statistically significant difference. 
			(c-e) Time courses of the species abundances simulated with system (1) corresponding to (a). In Fig.~\ref{Hudata}a: $r_{i}=0.4$, $d_{i}=0.001$, $\alpha_{ij}=\mathcal{N}(-0.1, 0.1)$.
			In Fig.~\ref{Hudata}a (Low):  $\beta_{ijk}=\mathcal{N}(-0.19, 0.24)$, $(i, j, k= 1,\cdots,20)$. In Fig.~\ref{Hudata}a (Medium): $\beta_{ijk}=\mathcal{N}(-0.3, 0.28)$, $(i, j, k= 1,\cdots,26)$. In Fig.~\ref{Hudata}a (Low): $\beta_{ijk}=\mathcal{N}(-0.32, 0.30)$, $(i, j, k= 1,\cdots,26)$.}   
	\end{figure}

	\begin{figure}[ht!]   
		\centering
		\includegraphics[width=16cm]{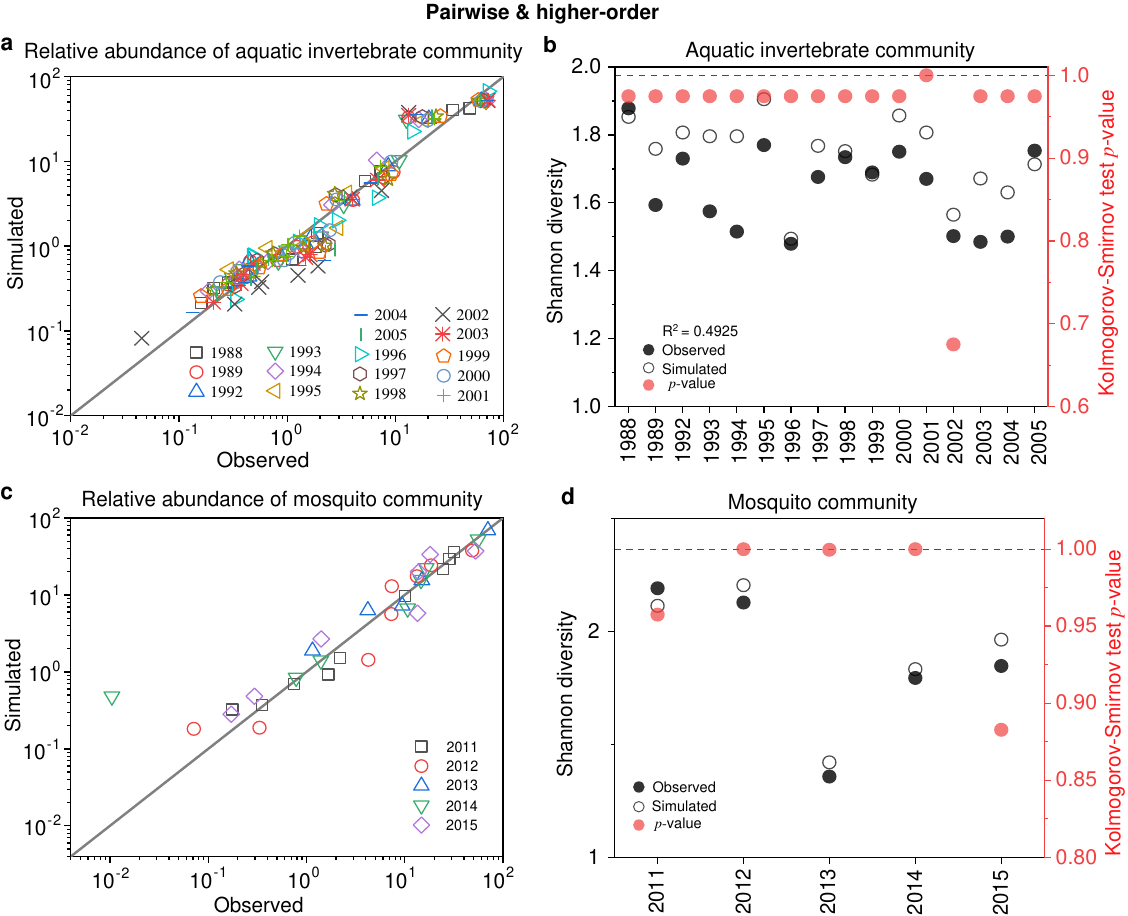}
		\caption{\label{NEE_mosquito-acquatic} Higher-order interaction illustrates the distribution pattern of the species’ in aquatic invertebrates and mosquitos community. (a, c) A visual comparison of species distribution across the aquatic invertebrates community from 1988-2005, and the mosquitos community from 2011-2015. These observed data reported in existing studies~\cite{MichaelSCrossley2020}, the simulated results were constructed from timestamp $ \textit{t} = 1.0\times10^{5} $ in the time series. 
			(b, d) The Shannon diversity and K-S test (\emph{p}-values) indicating whether the simulation results and the corresponding observed data come from identical distributions in (a), and in the K-S test, using a significance threshold of 0.05, none of the \emph{p}-values indicate a statistically significant difference. In Fig.~\ref{NEE_mosquito-acquatic}a, c:  $d_{i}=0.001$, $\alpha_{ij}=\mathcal{N}(-0.1, 0.1)$.
			In Fig.~\ref{NEE_mosquito-acquatic}a (1988): $r_{i}=0.4$, $\beta_{ijk}=\mathcal{N}(-0.28, 0.19)$, $(i, j, k= 1,\cdots,10)$. In Fig.~\ref{NEE_mosquito-acquatic}a (1989): $r_{i}=0.35$, $\beta_{ijk}=\mathcal{N}(-0.2, 0.2)$, $(i, j, k= 1,\cdots,10)$. In Fig.~\ref{NEE_mosquito-acquatic}a (1992): $r_{i}=0.35$, $\beta_{ijk}=\mathcal{N}(-0.2, 0.15)$, $(i, j, k= 1,\cdots,10)$. In Fig.~\ref{NEE_mosquito-acquatic}a (1993): $r_{i}=0.35$, $\beta_{ijk}=\mathcal{N}(-0.15, 0.1)$, $(i, j, k= 1,\cdots,10)$. In Fig.~\ref{NEE_mosquito-acquatic}a (1994): $r_{i}=0.35$, $\beta_{ijk}=\mathcal{N}(-0.15, 0.1)$, $(i, j, k= 1,\cdots,10)$. In Fig.~\ref{NEE_mosquito-acquatic}a (1995): $r_{i}=0.3$, $\beta_{ijk}=\mathcal{N}(-0.25, 0.2)$, $(i, j, k= 1,\cdots,10)$. In Fig.~\ref{NEE_mosquito-acquatic}a (1996): $r_{i}=0.25$, $\beta_{ijk}=\mathcal{N}(-0.25, 0.1)$, $(i, j, k= 1,\cdots,10)$. In Fig.~\ref{NEE_mosquito-acquatic}a (1997): $r_{i}=0.3$, $\beta_{ijk}=\mathcal{N}(-0.2, 0.3)$, $(i, j, k= 1,\cdots,10)$. In Fig.~\ref{NEE_mosquito-acquatic}a (1998): $r_{i}=0.3$, $\beta_{ijk}=\mathcal{N}(-0.2, 0.35)$, $(i, j, k= 1,\cdots,10)$. In Fig.~\ref{NEE_mosquito-acquatic}a (1999): $r_{i}=0.35$, $\beta_{ijk}=\mathcal{N}(-0.15, 0.2)$, $(i, j, k= 1,\cdots,10)$. In Fig.~\ref{NEE_mosquito-acquatic}a (2000): $r_{i}=0.35$, $\beta_{ijk}=\mathcal{N}(-0.2, 0.12)$, $(i, j, k= 1,\cdots,10)$. In Fig.~\ref{NEE_mosquito-acquatic}a (2001): $r_{i}=0.35$, $\beta_{ijk}=\mathcal{N}(-0.2, 0.15)$, $(i, j, k= 1,\cdots,10)$. In Fig.~\ref{NEE_mosquito-acquatic}a (2002): $r_{i}=0.5$, $\beta_{ijk}=\mathcal{N}(-0.15, 0.6)$, $(i, j, k= 1,\cdots,10)$. In Fig.~\ref{NEE_mosquito-acquatic}a (2003): $r_{i}=0.4$, $\beta_{ijk}=\mathcal{N}(-0.2, 0.3)$, $(i, j, k= 1,\cdots,10)$. In Fig.~\ref{NEE_mosquito-acquatic}a (2004): $r_{i}=0.45$, $\beta_{ijk}=\mathcal{N}(-0.2, 0.35)$, $(i, j, k= 1,\cdots,10)$. In Fig.~\ref{NEE_mosquito-acquatic}a (2005): $r_{i}=0.35$, $\beta_{ijk}=\mathcal{N}(-0.2, 0.3)$, $(i, j, k= 1,\cdots,10)$. In Fig.~\ref{NEE_mosquito-acquatic}c (2011): $r_{i}=0.25$, $\beta_{ijk}=\mathcal{N}(-0.2, 0.22)$, $(i, j, k= 1,\cdots,9)$. In Fig.~\ref{NEE_mosquito-acquatic}c (2012): $r_{i}=0.4$, $\beta_{ijk}=\mathcal{N}(-0.16, 0.4)$, $(i, j, k= 1,\cdots,8)$. In Fig.~\ref{NEE_mosquito-acquatic}c (2013): $r_{i}=0.2$, $\beta_{ijk}=\mathcal{N}(-0.25, 0.23)$, $(i, j, k= 1,\cdots,5)$. In Fig.~\ref{NEE_mosquito-acquatic}c (2014): $r_{i}=0.5$, $\beta_{ijk}=\mathcal{N}(-0.42, 0.45)$, $(i, j, k= 1,\cdots,7)$. In Fig.~\ref{NEE_mosquito-acquatic}c (2015): $r_{i}=0.25$, $\beta_{ijk}=\mathcal{N}(-0.2, 0.28)$, $(i, j, k= 1,\cdots,7)$.}   
	\end{figure}

	\begin{figure}[ht!]   
		\centering
		\includegraphics[width=16cm]{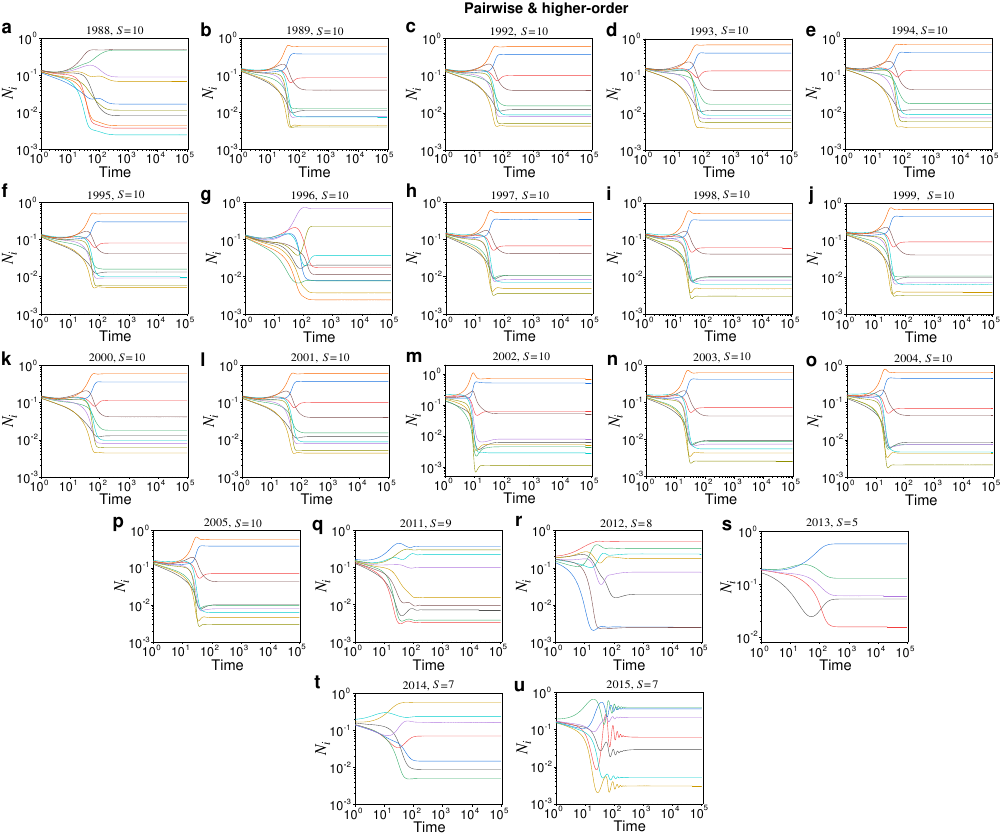}
		\caption{\label{NEE_mosquito-acquatic-timeseries} Higher-order interaction enables a wide range of consumer species to coexist. (a-u) Time courses of the species abundances simulated with system (1). The time series in (a-p) and (q-u) correspond to that shown in Fig.~\ref{NEE_mosquito-acquatic}a and Fig.~\ref{NEE_mosquito-acquatic}c, respectively. }   
	\end{figure}

	\begin{figure}[ht!]   
		\centering
		\includegraphics[width=16cm]{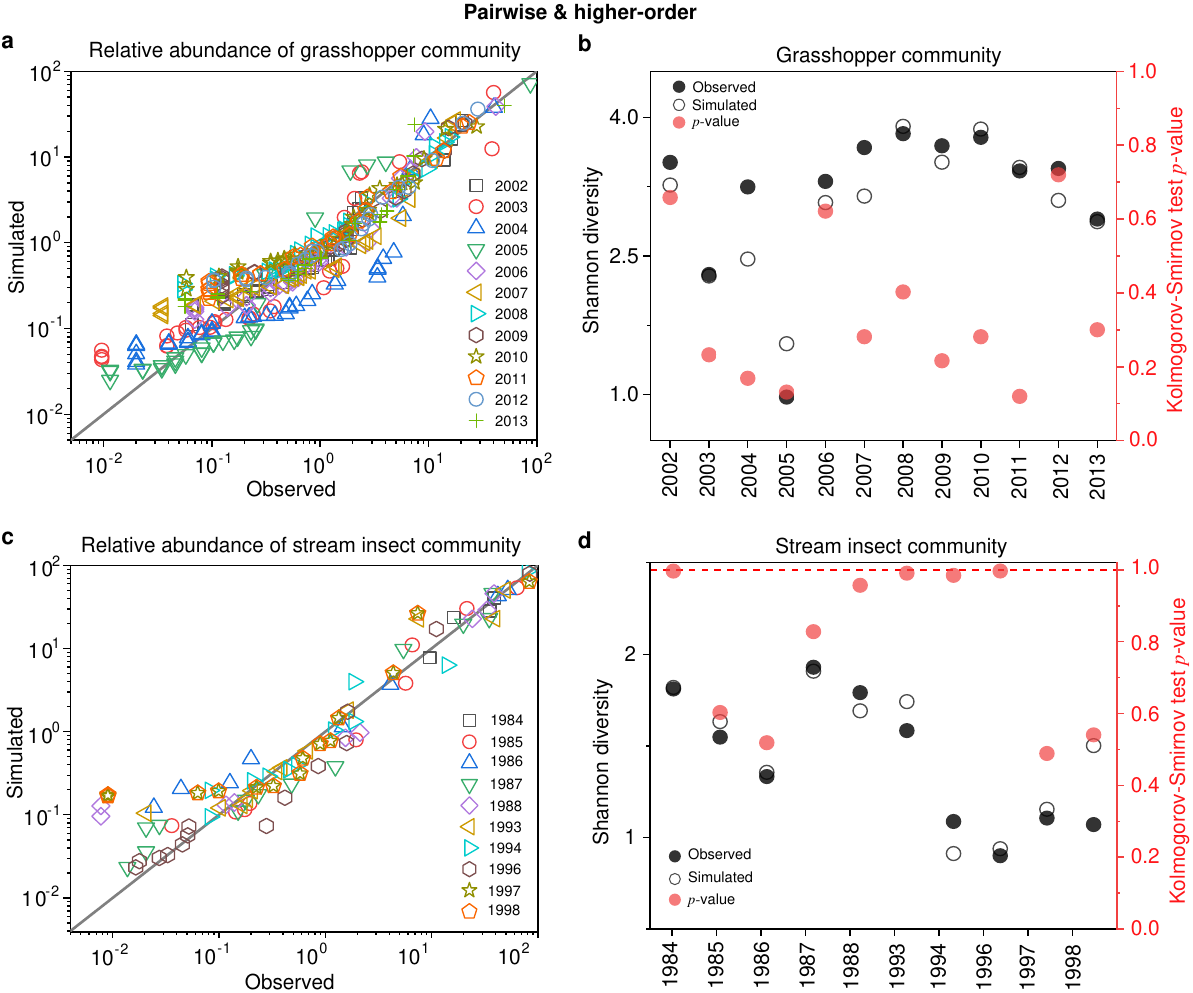}
		\caption{\label{NEE-Stream insects-Sevilleta} Higher-order interaction illustrates the distribution pattern of the species’ in stream insect and grasshopper communities. (a, c) A visual comparison of species distribution across the stream insect communities from 1984-1998, and the grasshopper community from 2002-2013. These observed data reported in existing studies~\cite{MichaelSCrossley2020}, the simulated results were constructed from timestamp $ \textit{t} = 1.0\times10^{5} $ in the time series. 
		(b, d) The Shannon diversity and K-S test (\emph{p}-values) indicating whether the simulation results and the corresponding observed data come from identical distributions in (a), and in the K-S test, using a significance threshold of 0.05, none of the \emph{p}-values indicate a statistically significant difference. In Fig.~\ref{NEE-Stream insects-Sevilleta}a, c:  $d_{i}=0.001$, $\alpha_{ij}=\mathcal{N}(-0.1, 0.1)$.
		In Fig.~\ref{NEE-Stream insects-Sevilleta}a (2002): $r_{i}=0.4$, $\beta_{ijk}=\mathcal{N}(-0.25, 0.39)$, $(i, j, k= 1,\cdots,36)$. In Fig.~\ref{NEE-Stream insects-Sevilleta}a (2003): $r_{i}=0.5$, $\beta_{ijk}=\mathcal{N}(-0.25, 0.4)$, $(i, j, k= 1,\cdots,32)$. In Fig.~\ref{NEE-Stream insects-Sevilleta}a (2004): $r_{i}=0.4$, $\beta_{ijk}=\mathcal{N}(-0.22, 0.35)$, $(i, j, k= 1,\cdots,43)$. In Fig.~\ref{NEE-Stream insects-Sevilleta}a (2005): $r_{i}=0.5$, $\beta_{ijk}=\mathcal{N}(-0.25, 0.3)$, $(i, j, k= 1,\cdots,32)$. In Fig.~\ref{NEE-Stream insects-Sevilleta}a (2006): $r_{i}=0.31$, $\beta_{ijk}=\mathcal{N}(-0.15, 0.32)$, $(i, j, k= 1,\cdots,34)$. In Fig.~\ref{NEE-Stream insects-Sevilleta}a (2007): $r_{i}=0.43$, $\beta_{ijk}=\mathcal{N}(-0.15, 0.3)$, $(i, j, k= 1,\cdots,35)$. In Fig.~\ref{NEE-Stream insects-Sevilleta}a (2008): $r_{i}=0.37$, $\beta_{ijk}=\mathcal{N}(-0.2, 0.45)$, $(i, j, k= 1,\cdots,33)$. In Fig.~\ref{NEE-Stream insects-Sevilleta}a (2009): $r_{i}=0.5$, $\beta_{ijk}=\mathcal{N}(-0.3, 0.5)$, $(i, j, k= 1,\cdots,31)$. In Fig.~\ref{NEE-Stream insects-Sevilleta}a (2010): $r_{i}=0.45$, $\beta_{ijk}=\mathcal{N}(-0.35, 0.45)$, $(i, j, k= 1,\cdots,35)$. In Fig.~\ref{NEE-Stream insects-Sevilleta}a (2011): $r_{i}=0.5$, $\beta_{ijk}=\mathcal{N}(-0.27, 0.45)$, $(i, j, k= 1,\cdots,31)$. In Fig.~\ref{NEE-Stream insects-Sevilleta}a (2012): $r_{i}=0.3$, $\beta_{ijk}=\mathcal{N}(-0.27, 0.45)$, $(i, j, k= 1,\cdots,28)$. In Fig.~\ref{NEE-Stream insects-Sevilleta}a (2013): $r_{i}=0.4$, $\beta_{ijk}=\mathcal{N}(-0.25, 0.4)$, $(i, j, k= 1,\cdots,28)$. In Fig.~\ref{NEE-Stream insects-Sevilleta}c (1984): $r_{i}=0.25$, $\beta_{ijk}=\mathcal{N}(-0.3, 0.22)$, $(i, j, k= 1,\cdots,4)$. In Fig.~\ref{NEE-Stream insects-Sevilleta}c (1985): $r_{i}=0.25$, $\beta_{ijk}=\mathcal{N}(-0.2, 0.35$, $(i, j, k= 1,\cdots,9)$. In Fig.~\ref{NEE-Stream insects-Sevilleta}c (1986): $r_{i}=0.38$, $\beta_{ijk}=\mathcal{N}(-0.18, 0.23)$, $(i, j, k= 1,\cdots,8)$. In Fig.~\ref{NEE-Stream insects-Sevilleta}c (1987): $r_{i}=0.4$, $\beta_{ijk}=\mathcal{N}(-0.18, 0.3)$, $(i, j, k= 1,\cdots,13)$. In Fig.~\ref{NEE-Stream insects-Sevilleta}c (1988): $r_{i}=0.4$, $\beta_{ijk}=\mathcal{N}(-0.2, 0.3)$, $(i, j, k= 1,\cdots,9)$. In Fig.~\ref{NEE-Stream insects-Sevilleta}c (1993): $r_{i}=0.4$, $\beta_{ijk}=\mathcal{N}(-0.2, 0.3)$, $(i, j, k= 1,\cdots,12)$. In Fig.~\ref{NEE-Stream insects-Sevilleta}c (1994): $r_{i}=0.3$, $\beta_{ijk}=\mathcal{N}(-0.25, 0.28)$, $(i, j, k= 1,\cdots,11)$. In Fig.~\ref{NEE-Stream insects-Sevilleta}c (1996): $r_{i}=0.35$, $\beta_{ijk}=\mathcal{N}(-0.23, 0.3)$, $(i, j, k= 1,\cdots,14)$. In Fig.~\ref{NEE-Stream insects-Sevilleta}c (1997): $r_{i}=0.4$, $\beta_{ijk}=\mathcal{N}(-0.2, 0.1)$, $(i, j, k= 1,\cdots,13)$. In Fig.~\ref{NEE-Stream insects-Sevilleta}c (1998): $r_{i}=0.32$, $\beta_{ijk}=\mathcal{N}(-0.2, 0.26)$, $(i, j, k= 1,\cdots,14)$.}   
	\end{figure}
	
	\begin{figure}[ht!]   
		\centering
		\includegraphics[width=16cm]{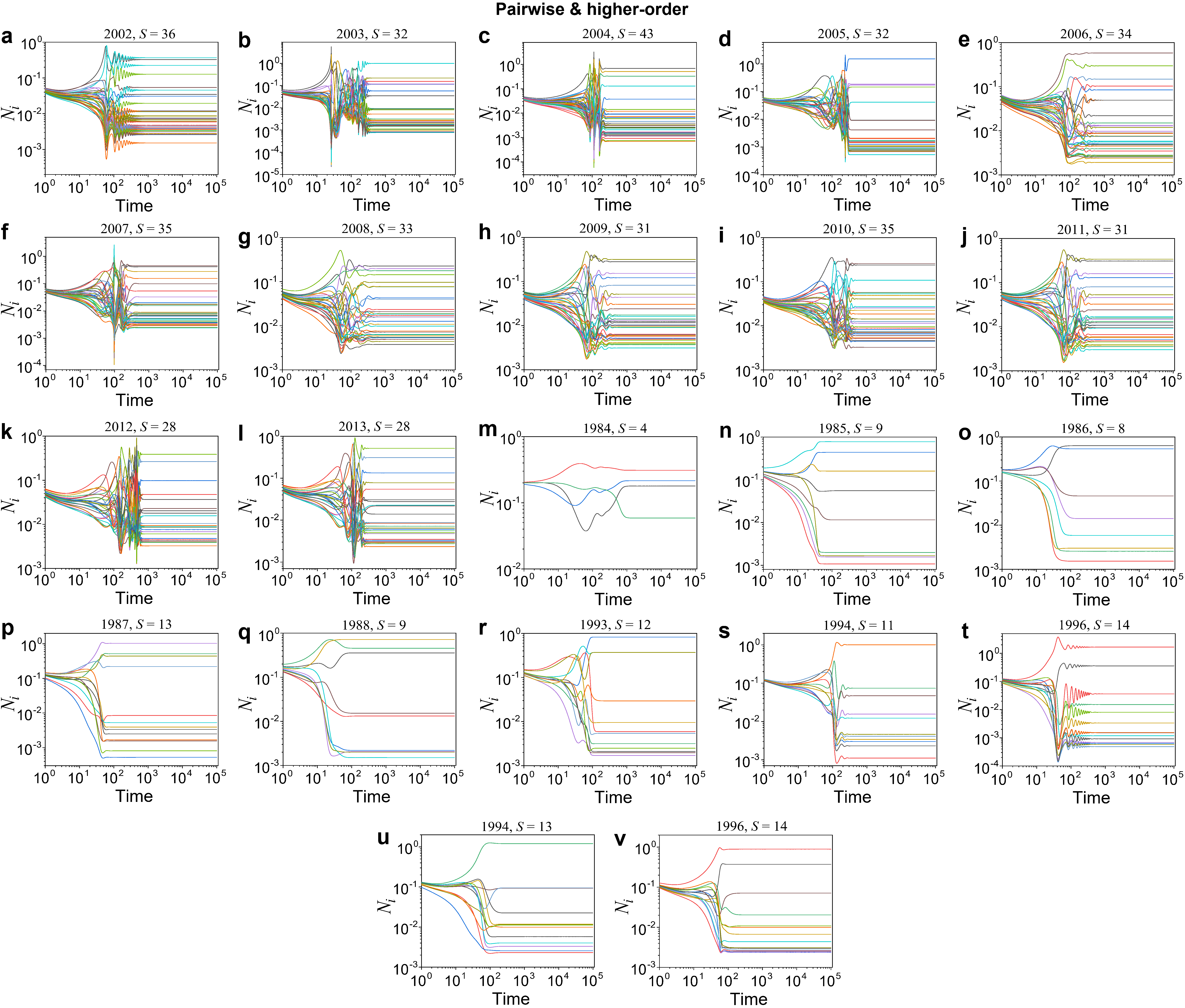}
		\caption{\label{NEE-Stream insects-Sevilleta-timeseries} Higher-order interaction enables a wide range of consumer species to coexist. (a-v) Time courses of the species abundances simulated with system (1). The time series in (a-l) and (m-v) correspond to that shown in Fig.~\ref{NEE-Stream insects-Sevilleta}a and Fig.~\ref{NEE-Stream insects-Sevilleta}c, respectively. }   
	\end{figure}	
		

\end{document}